\documentclass[onecolumn]{IEEEtran}
\usepackage{mathtools,amsthm,amssymb,amsfonts,bm}
\usepackage{thmtools,amsmath,extarrows}
\usepackage{algorithmic}
\usepackage{algorithm}
\usepackage{array,tabularx,booktabs,multirow,colortbl}
\usepackage{makecell}
\usepackage{tikz-cd,physics}
\usepackage[all]{xy}
\usepackage[caption=false,font=normalsize,labelfont=sf,textfont=sf]{subfig}
\usepackage{textcomp}
\usepackage{stfloats}
\usepackage{url}
\usepackage{verbatim}
\usepackage{graphicx}
\usepackage[noadjust]{cite}
\newtheorem{definition}{Definition}
\newtheorem{lemma}{Lemma}
\newtheorem{theorem}{Theorem}
\newtheorem{construction}{Construction}
\newtheorem{remark}{Remark}

\newtheorem{corollary}{Corollary}
\newtheorem{example}{Example}
\newcommand{\unchanged}{\mathcal{U}}
\newcommand{\reading}{\mathcal{R}}
\newcommand{\written}{\mathcal{W}}

\begin{document}

\title{MDS Generalized Convertible Code}

\author{Songping Ge, Han Cai, \IEEEmembership{Member,~IEEE}, and Xiaohu Tang, \IEEEmembership{Senior Member,~IEEE}
\thanks{S. Ge, H. Cai and X. Tang are with the School of Information Science and Technology,
  Southwest Jiaotong University, Chengdu, 610031, China (e-mail: gesongping@my.swjtu.edu.cn; hancai@aliyun.com; xhutang@swjtu.edu.cn).}
}

\maketitle

\begin{abstract}
In this paper, we consider the convertible codes with the maximum distance separable (MDS) property, which can adjust the code rate according to the failure rates of devices. We first extend the notion of convertible codes to allow initial and final codes with different parameters. Then, we investigate the relationship between these parameters and thus establish new lower bounds on the access cost in the merge and split regimes. To gain a deeper understanding of access-optimal MDS convertible codes in the merge regime, we characterize them from the perspective of parity check matrices. Consequently, we present a necessary and sufficient condition for the access-optimal MDS convertible code in the merge regime. Finally, as an application of our characterization, we construct MDS convertible codes in the merge regime with optimal access cost based on the extended generalized Reed-Solomon codes.
\end{abstract}


\section{Introduction}
Large-scale cloud storage and distributed file systems, such as Amazon Elastic
Block Store (EBS) and Google File System (GoogleFS), have grown so immense that disk failures are more common than exceptional. Thus, we must properly impose redundancy in these systems to guarantee data reliability. One
possible solution is to apply erasure coding to protect against failures. A set of $k$ original information symbols is encoded into an $n$-length vector
as a codeword employing an $[n, k]$ code and stored across $n$ storage devices (nodes).
In this system, $k/n$ is known as the code rate. Typically, the code rate is fixed, despite the evolving failure rates of storage devices over time, which will result in either the inability to ensure data reliability during periods of high erasure rates or excessively low coding rates and resource wastage during periods of low erasure rates. To address this
issue, Kadekodi, Rashmi, and Ganger \cite{Kadekodi2019} proposed an economically viable
solution, advocating for flexibility in adjusting redundancy levels over time.
Their approach involves converting ability for codes that can convert codewords
 of an $[n_I, k_I]$ initial code to final code with different parameters $[n_F, k_F]$
 while retaining the original message. This process, termed \textit{code conversion}
 \cite{Maturana2020a},  demonstrates a reduction
 in storage space ranging from $11\%$ to $44\%$ as shown in \cite{Kadekodi2019}.

In general, the code conversion is to decode the original message from the \textit{initial code} and then re-encode it into the \textit{final code}, which needs to access a large number of symbols to rebuild the final codeword. There are two special cases for code conversion: On one hand, multiple initial codewords are merged into one final codeword, called the \textit{merge regime}. On the other hand, one initial codeword is split into multiple final codewords, called the \textit{split regime}. For convertible codes, one of the most important parameters is the access cost, which counts the number of symbols accessed by reading the original codewords and the number of symbols written in the final codewords. However, the access cost is not independent of the other parameters such as code lengths, dimensions, and code rates. In \cite{Maturana2022} and \cite{Maturana2020}, the theoretic bounds on these parameters were derived, and then explicit constructions of MDS convertible codes with optimal access cost were presented. Very recently, Kong \cite{Kong2023} introduced a construction of MDS convertible codes with optimal access cost in the merge regime, which may have a linear field size compared with the final code length. In addition, \cite{Kong2023} and \cite{Maturana2023} focused on the convertible codes with locality, and \cite{Maturana2023a} studied the fundamental limits on bandwidth during code conversion and proposed bandwidth-optimal convertible codes.

Nonetheless, all the known results consider two regular cases. That is, all the initial codes in the merge regime or final codes in the split regime share the same parameters:  code length and dimension, or redundancy. In distributed storage systems, this may lead to a uniform constraint on all the nodes, such as having identical failure rates for all storage devices.
 To remove this constraint, we first expand the definition of convertible codes, which allows initial codes and final codes to have different parameters.
 We then focus on the conversion problem for MDS codes under this general setting and investigate the relationship between parameters by establishing new lower bounds on the access cost for the convertible codes in the merge and split regimes. According to these lower bounds, the optimal access cost of convertible codes in the split regime is only related to the parameters of one final code. Constructing convertible codes with optimal access cost in the split regime is therefore relatively straightforward. In this paper we considers constructions of convertible codes in the merge regime. To better understand the convertible codes with optimal access cost in the merge regime, we characterize convertible codes via parity check matrices. As a result, we present a necessary and sufficient condition for convertible codes in the merge regime to achieve the lower bound on the access cost with equality, which may guide the construction of convertible codes with optimal access cost. At last, as an application of our characterization, we present a construction of convertible codes in the merge regime with optimal access cost via the extended generalized Reed-Solomon codes. Notably, our construction not only covers the known one by Kong \cite{Kong2023} but also relaxes any restriction of other parameters compared to \cite{Maturana2022} and \cite{Kong2023}. Moreover, it yields new codes having the theoretical minimum filed size due to the MDS-conjecture \cite{Huffman2003}. For comparison, the parameters of the known access-optimal MDS convertible codes in the merge regime, as well as a new one of this paper, are listed in Table \ref{compare}.

The subsequent sections of the paper are organized as follows: Section \ref{sec_preliminary} reviews notation and preliminaries. Section \ref{sec-generalized_CC} introduces the definition of generalized convertible codes. Section \ref{sec_bounds} presents new lower bounds on access cost for MDS generalized convertible codes in the merge and split regimes. Section \ref{sec-structure} characterizes the access-optimal  MDS generalized convertible codes in the merge regime. Section \ref{sec_cons} proposes a construction of access-optimal  MDS generalized convertible codes in the merge regime. Finally, Section \ref{sec_conclusion} concludes this paper.

\renewcommand{\arraystretch}{1.5}
\begin{table}[t!]
    \centering
    \caption{Known $(t,1)_q$ access-optimal linear MDS merge-convertible Code where each initial code $\mathcal{C}_{I_i}$ is $[n_{I_i},k_{I_i}]_q$ MDS codes for $i\in[t]$, and final code $\mathcal{C}_{F}$ is $[n_{F},k_{F}]$ MDS code. Denote $r_{I_i}\triangleq n_{I_i}-k_{I_i}$ and $r_{F}\triangleq n_{F}-k_{F}$.}
    \begin{tabular}{|c|c|c|c|}
    \hline
    \textbf{Construction} & \textbf{Restrictions}  & \textbf{Field size $q$ requirement} & \textbf{Reference} \\
    \hline
    \hline
    \multirow{2}{*}{General} & \multirow{2}{*}{\makecell{$\mathcal{C}_{I_1}=\mathcal{C}_{I_2}=\cdots=\mathcal{C}_{I_t}$,\\ $r_F\le\min\{k_{I_1},r_{I_1}\}$}} &  \multirow{2}{*}{$\max\left\{ \makecell{2^{O((n_F)^3)},\\ n_{I_1}-1} \right\}$} & \multirow{2}{*}{\cite[Theorem 21]{Maturana2022}}  \\
    & & &\\
    \hline
    \multirow{2}{*}{Hankel-I} & \multirow{2}{*}{\makecell{$\mathcal{C}_{I_1}=\mathcal{C}_{I_2}=\cdots=\mathcal{C}_{I_t}$,\\ $r_F\le \lfloor r_{I_1}/t\rfloor$}} &  \multirow{2}{*}{$\max\{n_F,n_{I_1}\}-1$} &  \multirow{2}{*}{\cite[Example 22]{Maturana2022}} \\
     & & &\\
    \hline
     \multirow{2}{*}{Hankel-II} & \multirow{2}{*}{\makecell{$\mathcal{C}_{I_1}=\mathcal{C}_{I_2}=\cdots=\mathcal{C}_{I_t}$,\\ $r_F\le r_{I_1}-t+1$}} &  \multirow{2}{*}{$\max\{ k_{I_1}r_{I_1},n_{I_1}-1\}$} & \multirow{2}{*}{\cite[Example 23]{Maturana2022}}  \\
     & & &\\
    \hline
     \multirow{4}{*}{\makecell{Hankel$_s$ for\\ $t\le s\le r_{I_1}$}} & \multirow{4}{*}{\makecell{$\mathcal{C}_{I_1}=\mathcal{C}_{I_2}=\cdots=\mathcal{C}_{I_t}$,\\ $r_F\le(s-t+1)\lfloor r_{I_1}/s \rfloor$\\ $+\max\{(r_{I_1}\mod s)-t+1,0\}$}} & \multirow{4}{*}{$\max\left\{ \makecell{sk_{I_1}+\lfloor r_{I_1}/s \rfloor-1,\\ n_{I_1}-1} \right\}$} & \multirow{4}{*}{\cite[Theorem 24]{Maturana2022}} \\
    & & &\\
    & & &\\
    & & &\\
    \hline
     \multirow{2}{*}{GRS} & \multirow{2}{*}{\makecell{$\mathcal{C}_{I_1}=\mathcal{C}_{I_2}=\cdots=\mathcal{C}_{I_t}$,\\ $r_{F}\le\min\{k_{I_1},r_{I_1}\}$}} & \multirow{2}{*}{\makecell{$(t+1)\max\{k_{I_1},r_{I_1}\}+1$,\\ $\max\{k_{I_1},r_{I_1}\}\mid(q-1)$}} & \multirow{2}{*}{\cite[Corollary II.2]{Kong2023}}  \\
     & & &\\
    \hline
    Extended GRS & None & $\max\{n_{I_1},\cdots,n_{I_{t}},n_{F_1}\}-1$ & Theorem \ref{thm_4}  \\
    \hline
    \end{tabular}
    \label{compare}
\end{table}

\section{Preliminaries}\label{sec_preliminary}
We first introduce some notation used throughout this paper.
\begin{itemize}
\item For a positive integer $a$, let $[a]$ be the set $\{1,2,\cdots,a\}$;
\item For a prime power $q$, let $\mathbb{F}_q$ be the finite field with $q$ elements and $\mathbb{F}_q^* = \mathbb{F}_q\setminus\{0\}$;
\item For two vector spaces $\mathbb{V}_1,\mathbb{V}_2$ over $\mathbb{F}_q$, denote $\mathbb{V}_1\otimes\mathbb{V}_2=\{(\bm{v_1},\bm{v_2}):\bm{v_1}\in\mathbb{V}_1,\bm{v_2}\in\mathbb{V}_2\}$;
\item For a finite set $\mathcal{S}$, let $|\mathcal{S}|$ denote its cardinality;
\item For a $m\times n$ matrix $M$ with columns indexed by $\mathcal{V}$ , and a set $\mathcal{B}\subseteq\mathcal{V}$, denote $M|_{\mathcal{B}}$ as the $m \times |\mathcal{B}|$ submatrix of $M$ with columns indexed by $\mathcal{B}$, where we keep the relative order of columns;
\item Let $\top$ be the transpose operator.
\end{itemize}

An $[n,k]_q$ linear code $\mathcal{C}$ over $\mathbb{F}_q$ is a $k$-dimensional subspace of $ \mathbb{F}_q^n$. In particular, if the minimum Hamming distance of $\mathcal{C}$ is $d=n-k+1$, then it is called an MDS code.
For simplicity, throughout this paper, we define $r\triangleq n-k$, which is the number of redundancies of code $\mathcal{C}$.

The \emph{parity check matrix} of  $\mathcal{C}$ is an $(n-k) \times n$ matrix with full rank  over $\mathbb{F}_q$ denoted by $H$, i.e., $\mathcal{C}=\{\bm{c}:\bm{c}\cdot H^{\top}=\bm{0}\}$. The \emph{generator matrix} $G$ of $\mathcal{C}$ is a $k\times n$ matrix with full rank over $\mathbb{F}_q$, i.e., $\mathcal{C}=\{\bm{m}\cdot G:\bm{m}\in\mathbb{F}_q^k\}$ where $\bm{m}$ is a message vector. Denote  $\mathcal{C}^\bot$ as  dual code of the code $\mathcal{C}$, where $\mathcal{C}^\bot$ has generator matrix $H$.

For a set $\mathcal{T}\subseteq [n]$, let $\mathcal{C}|_\mathcal{T}$ denote the punctured code of  $\mathcal{C}$  formed by all the codewords of $\mathcal{C}$ that restricts the symbols in coordinate set $\mathcal{T}$. For a vector $\bm{x}$, denote $\bm{x}|_{\mathcal{T}}$ as the restriction of $\bm{x}$ onto the coordinates set $\mathcal{T}$.

\begin{lemma}[{\cite{Huffman2003}}]\label{lem: prop of puc and srt}
    Let $\mathcal{C}$ be an $[n,k]_q$ MDS code. Then $\mathcal{C}|_\mathcal{T}$ is a $[|\mathcal{T}|,k]_q$ MDS code for arbitrary set $\mathcal{T}$ with $|\mathcal{T}|>k$.
\end{lemma}

\begin{definition}\label{def_generated}
Let $\mathcal{C}$ and $\mathcal{C}'$ be $[n,k]_q$ and $[n',k']_q$  codes, respectively. We say that code symbols of $\mathcal{C}$ can be linearly generated by code symbols of $\mathcal{C}'$, if there exists a $k\times k'$ matrix $U$ and a $n'\times n$ matrix $V$ over $\mathbb{F}_q$ such that
\begin{equation}\label{eqn_generate}
    G=UG'V,
\end{equation}
where $G$ and $G'$ are generator matrices of $\mathcal{C}$ and $\mathcal{C}'$, respectively.
\end{definition}

\begin{definition}\label{def_vand}
For two positive integers $n,r$ with $n>r$ and two vectors $\gamma=(\gamma_1,\gamma_2,\cdots,\gamma_{n-1}) \in \mathbb{F}_{q}^{n-1}$, $\mathbf{w}=(w_1,\cdots,w_{n-1},w_n)\in(\mathbb{F}_q^*)^n$, the extended Vandermond-type matrix is defined as
\begin{eqnarray*}
    \mathbf{Vand}_{r,n}(\gamma,\mathbf{w})\triangleq
     \begin{pmatrix}
        w_1 & w_2 & \cdots  & w_{n-1} & 0 \\
        w_1\gamma_1 & w_2\gamma_2 & \cdots & w_{n-1}\gamma_{n-1} & 0 \\
        \vdots & \vdots & \ddots  & \vdots & \vdots\\
        w_1\gamma_1^{{r-2}} & w_2\gamma_2^{{r-2}} & \cdots & w_{n-1}\gamma_{n-1}^{{r-2}} & 0\\
        w_1\gamma_1^{{r-1}} & w_2\gamma_2^{{r-1}} & \cdots & w_{n-1}\gamma_{n-1}^{{r-1}} & w_n
      \end{pmatrix}.
\end{eqnarray*}
\end{definition}

It is well known that any $\lambda$ columns from $\mathbf{Vand}_{r,n}(\gamma,\mathbf{w})$ are independent over $\mathbb{F}_q$ if $\lambda\le r$ and
$\gamma_i\ne \gamma_j,1\le i\ne j\le {n-1}$.  Therefore, based on  $\mathbf{Vand}_{r,n}(\gamma,\mathbf{w})$, we can obtain
the so called \textit{extended generalized Reed-Solomon (extended GRS) code} in \cite{Huffman2003}, which is an important class of MDS codes.

\begin{lemma}[{\cite{Huffman2003}}]\label{lem_GRS}
For two positive integers $r$ and $n$ with $r<n$. Let $\mathbf{\gamma}=(\gamma_1,\cdots,\gamma_{n-1})$ contain $n-1$ distinct elements of $\mathbb{F}_q$ and $\mathbf{w}=(w_1,\cdots,w_{n-1},w_n)\in(\mathbb{F}_q^*)^n$.
Then, the code $\mathcal{C}$  with the parity check matrix $\mathbf{Vand}_{r,n}(\gamma,\mathbf{w})$ is an
$[n,n-r]_q$ MDS code. Further, for any set
$\mathcal{T}\subseteq[n]$ with $\{n\}\subseteq\mathcal{T}$ and $|\mathcal{T}|>n-r$,  the punctured code $\mathcal{C}|_{\mathcal{T}}$
is a $[|\mathcal{T}|,n-r]_q$ MDS code having parity check matrix $\mathbf{Vand}_{r-(n-|\mathcal{T}|),|\mathcal{T}|}(\gamma|_{\mathcal{T}\setminus\{n\}},\theta_{\mathcal{T}}(\mathbf{w}))$ for a vector $\theta_{\mathcal{T}}(\mathbf{w})=(w_1^*,\cdots, w_{\mathcal{T}}^* )\in(\mathbb{F}_q^*)^{|\mathcal{T}|}$ determined by $\mathbf{w}$.
\end{lemma}

\section{Generalized Convertible Codes}\label{sec-generalized_CC}
In this section, we present a more general definition of convertible codes compared with the ones in \cite{Maturana2020a} and \cite{Maturana2020}, where the initial codes in the merge regime or final codes in the split regime are the same or have the same redundancy.

\begin{definition}[Generalized Convertible Code]\label{def-cc}
Let $t_1,t_2$ be two positive integers. A $(t_1,t_2)_q$ generalized convertible code $\bm{\mathcal{C}}$ over $\mathbb{F}_q$  consists of:

     1) $t_1$ initial codes $\mathcal{C}_{I_1},\mathcal{C}_{I_2},\cdots,\mathcal{C}_{I_{t_1}}$ and
     $t_2$ final codes $\mathcal{C}_{F_1},\mathcal{C}_{F_2},\cdots,\mathcal{C}_{F_{t_2}}$, where $\mathcal{C}_{I_i}$ is an $[n_{I_i},k_{I_i}]_q$ code with $r_{I_i}\triangleq n_{I_i}-k_{I_i}$ for $i\in[t_1]$,  and $\mathcal{C}_{F_j}$ is an $[n_{F_j},k_{F_j}]_q$ code with $r_{F_j}\triangleq  n_{F_j}-k_{F_j}$ for $j\in[t_2]$, such that
     \begin{equation}\label{equ:dimension}
        \sum_{i\in[t_1]}k_{I_i}=\sum_{j\in[t_2]}k_{F_j};
     \end{equation}

     2) A conversion procedure is determined by a  function $\sigma = (\sigma_1,\sigma_2,\cdots,\sigma_{t_2})$  from $\bigotimes_{i\in [t_1]}\mathcal{C}_{I_{i}} \rightarrow \bigotimes_{j\in [t_2]}\mathcal{C}_{F_j}$, i.e.,
    \begin{equation*}
        \sigma_j: \bigotimes_{i\in[t_1]}\mathcal{C}_{I_{i}} \rightarrow \mathcal{C}_{F_j},\,1\le j\le t_2.
    \end{equation*}

Specifically, a concrete conversion procedure is realized in three steps.

\textbf{Step 1: Choosing unchanged symbols from initial codes.}  For $j\in [t_2]$, the conversion procedure keeps some code symbols from each initial code $\mathcal{C}_{I_i},1\le i\le t_1$. Denote the indices of these symbols as  $\unchanged_{1,j},\unchanged_{2,j},\cdots,\unchanged_{t_1,j}$, where $\unchanged_{i,j}\subseteq\{i\}\times[n_{I_i}]$ satisfies
 \begin{equation*}
 \unchanged_{i,\tau_1}\cap \unchanged_{i,\tau_2}=\emptyset \,\, \text{ for }\tau_1\ne \tau_2\in[t_2].
\end{equation*}
For the convenience of analysis, we use two-dimensional indices to
denote the code symbols, i.e., the codeword $\bm{c}_{I_i}=(c_{i,1},c_{i,2},$ $\cdots, c_{i,n_{I_i}})\in \mathcal{C}_{I_i}$ for $i\in [t_1]$.

\textbf{Step 2: Reading symbols from  initial codes.} For $j\in [t_2]$, the function $\sigma_j$ reads some code symbols from each initial code $\mathcal{C}_{I_i},1\le i\le t_1$, whose indices are denoted by $\reading_{1,j},\reading_{2,j},\cdots,\reading_{t_1,j}$ where $\reading_{i,j}\subseteq\{i\}\times[n_{I_{i}}]$, respectively.

\textbf{Step 3: Conversion.} For any $j\in [t_2]$, choose $u_j$ unchanged symbols $\mathcal{C}_{I_1}|_{\unchanged_{1,j}},\mathcal{C}_{I_2}|_{\unchanged_{2,j}},\cdots,\mathcal{C}_{I_{t_1}}|_{\unchanged_{t_1,j}}$. And then  apply  $\sigma_j$ to the read symbols $\mathcal{C}_{I_1}|_{\reading_{1,j}},\mathcal{C}_{I_2}|_{\reading_{2,j}},\cdots,\mathcal{C}_{I_{t_1}}|_{\reading_{t_1,j}}$ to generate
$n_{F_j}-u_j$ symbols
\begin{equation}\label{Eqn_Gen_Fj}
\mathcal{C}_{F_j}|_{\written_{j}}
= \left\{ \sigma_j(\bm{x}) : \bm{x}\in\bigotimes_{i\in[t_1]}\mathcal{C}_{I_i}|_{\reading_{i,j}} \right\},
\end{equation}
 which are called written symbols, where $\written_{j}\triangleq \{w_j\}\times[n_{F_j}-u_j]$ and $w_j\triangleq t_1+j$.

Finally, the final code $\mathcal{C}_{F_j}$ is given by
\begin{align}
\mathcal{C}_{F_j}
&=\left\{ (\bm{x}_1,\sigma_j(\bm{x}_2)) : \bm{x}\in\bigotimes_{i\in[t_1]}\mathcal{C}_{I_i}, \bm{x}_1=\bm{x}|_{\bigcup_{i\in[t_1]}\unchanged_{i,j}}, \bm{x}_2=\bm{x}|_{\bigcup_{i\in[t_1]}\reading_{i,j}} \right\} \label{eqn_5}\\
&\subseteq \bigotimes_{i\in[t_1]}\mathcal{C}_{I_i}|_{\unchanged_{i,j}} \otimes \mathcal{C}_{F_j}|_{\written_j} \notag
\end{align}
for each $1\le j\le t_2$, where we assume the code symbols of $\mathcal{C}_{F_j}$ are
indexed by the set $$\bigcup_{i\in[t_1]}\unchanged_{i,j}\cup\written_{j}$$
such that
\begin{equation}\label{eqn_1}
    \mathcal{C}_{I_i}|_{\unchanged_{i,j}} = \mathcal{C}_{F_j}|_{\unchanged_{i,j}}.
\end{equation}

In particular, $\bm{\mathcal{C}}$ is said to be an  MDS generalized convertible code if all the initial and final codes are  MDS codes, and $\bm{\mathcal{C}}$ is called linear convertible code if all conversion procedure functions are linearly defined by Definition \ref{def_generated}.
\end{definition}

In this paper, we focus on linear generalized convertible code with linear initial and final codes. Therefore, in the sequel all the codes and conversion procedure functions are linear unless otherwise stated. For a  $(t_1,t_2)_q$ generalized convertible code $\bm{\mathcal{C}}$,  there are two special regimes, i.e., merge and split regimes, which are of particular interest.

\begin{definition}\label{def_MC_stable}
Let $\bm{\mathcal{C}}$ be a $(t_1,t_2)_q$ generalized convertible code. The convertible code $\bm{\mathcal{C}}$ is said to be
\begin{enumerate}
    \item a generalized merge-convertible code if $t_1>1$ and $t_2=1$; or
    \item a generalized split-convertible code if $t_1=1$ and $t_2>1$.
\end{enumerate}
\end{definition}

\begin{remark}
The generalized convertible codes expand the parameter range of the conventional convertible codes defined by Maturana and Rashmi, which only includes the merge and split regimes. Particularly,  when $t_1>1,t_2=1$, and $\mathcal{C}_{I_1}=\mathcal{C}_{I_2}=\cdots=\mathcal{C}_{I_{t_1}}$ (All the initial codes are the same) or $r_{I_1}=r_{I_2}=\cdots=r_{I_{t_1}}$ (All the initial codes have the same redundancy), it is the conventional merge code defined in \cite{Maturana2020a,Maturana2020}; While, when $t_1=1,t_2>1$, and $\mathcal{C}_{F_1}=\mathcal{C}_{F_2}=\cdots=\mathcal{C}_{F_{t_2}}$ (All the final codes are the same) or $r_{F_1}=r_{F_2}=\cdots=r_{F_{t_2}}$ (All the final codes have the same redundancy), it is the conventional split code defined in \cite{Maturana2020}.
\end{remark}

\begin{figure}[!t]
\centering
\tikzset {_pjc13y4it/.code = {\pgfsetadditionalshadetransform{ \pgftransformshift{\pgfpoint{0 bp } { 0 bp }  }  \pgftransformrotate{-45 }  \pgftransformscale{2 }  }}}
\pgfdeclarehorizontalshading{_3k5a8hzom}{150bp}{rgb(0bp)=(0.78,0.87,0.93);
rgb(37.5bp)=(0.78,0.87,0.93);
rgb(49.75bp)=(0.78,0.87,0.93);
rgb(50bp)=(0.98,0.95,0.61);
rgb(62.5bp)=(0.98,0.95,0.61);
rgb(100bp)=(0.98,0.95,0.61)}
\tikzset{_251ab1jdm/.code = {\pgfsetadditionalshadetransform{\pgftransformshift{\pgfpoint{0 bp } { 0 bp }  }  \pgftransformrotate{-45 }  \pgftransformscale{2 } }}}
\pgfdeclarehorizontalshading{_s5rbpw5oq} {150bp} {color(0bp)=(transparent!0);
color(37.5bp)=(transparent!0);
color(49.75bp)=(transparent!0);
color(50bp)=(transparent!25);
color(62.5bp)=(transparent!25);
color(100bp)=(transparent!25) }
\pgfdeclarefading{_9ypxvc9ks}{\tikz \fill[shading=_s5rbpw5oq,_251ab1jdm] (0,0) rectangle (50bp,50bp); }


\tikzset {_lzgwlebce/.code = {\pgfsetadditionalshadetransform{ \pgftransformshift{\pgfpoint{0 bp } { 0 bp }  }  \pgftransformrotate{-45 }  \pgftransformscale{2 }  }}}
\pgfdeclarehorizontalshading{_7os5ngkt3}{150bp}{rgb(0bp)=(0.78,0.87,0.93);
rgb(37.5bp)=(0.78,0.87,0.93);
rgb(49.75bp)=(0.78,0.87,0.93);
rgb(50bp)=(0.98,0.95,0.61);
rgb(62.5bp)=(0.98,0.95,0.61);
rgb(100bp)=(0.98,0.95,0.61)}
\tikzset{_tjmre60sh/.code = {\pgfsetadditionalshadetransform{\pgftransformshift{\pgfpoint{0 bp } { 0 bp }  }  \pgftransformrotate{-45 }  \pgftransformscale{2 } }}}
\pgfdeclarehorizontalshading{_qmxhbai7y} {150bp} {color(0bp)=(transparent!0);
color(37.5bp)=(transparent!0);
color(49.75bp)=(transparent!0);
color(50bp)=(transparent!25);
color(62.5bp)=(transparent!25);
color(100bp)=(transparent!25) }
\pgfdeclarefading{_6g4v1qgqe}{\tikz \fill[shading=_qmxhbai7y,_tjmre60sh] (0,0) rectangle (50bp,50bp); }


\tikzset {_hodct11gz/.code = {\pgfsetadditionalshadetransform{ \pgftransformshift{\pgfpoint{0 bp } { 0 bp }  }  \pgftransformrotate{-45 }  \pgftransformscale{2 }  }}}
\pgfdeclarehorizontalshading{_7duo1kxfb}{150bp}{rgb(0bp)=(0.78,0.87,0.93);
rgb(37.5bp)=(0.78,0.87,0.93);
rgb(49.75bp)=(0.78,0.87,0.93);
rgb(50bp)=(0.98,0.95,0.61);
rgb(62.5bp)=(0.98,0.95,0.61);
rgb(100bp)=(0.98,0.95,0.61)}
\tikzset{_3vyof9gyi/.code = {\pgfsetadditionalshadetransform{\pgftransformshift{\pgfpoint{0 bp } { 0 bp }  }  \pgftransformrotate{-45 }  \pgftransformscale{2 } }}}
\pgfdeclarehorizontalshading{_i2uz4zs18} {150bp} {color(0bp)=(transparent!0);
color(37.5bp)=(transparent!0);
color(49.75bp)=(transparent!0);
color(50bp)=(transparent!25);
color(62.5bp)=(transparent!25);
color(100bp)=(transparent!25) }
\pgfdeclarefading{_po9bf0tx0}{\tikz \fill[shading=_i2uz4zs18,_3vyof9gyi] (0,0) rectangle (50bp,50bp); }


\tikzset {_uae9mdzdk/.code = {\pgfsetadditionalshadetransform{ \pgftransformshift{\pgfpoint{0 bp } { 0 bp }  }  \pgftransformrotate{-45 }  \pgftransformscale{2 }  }}}
\pgfdeclarehorizontalshading{_pfd3bzl2l}{150bp}{rgb(0bp)=(0.78,0.87,0.93);
rgb(37.5bp)=(0.78,0.87,0.93);
rgb(49.75bp)=(0.78,0.87,0.93);
rgb(50bp)=(0.98,0.95,0.61);
rgb(62.5bp)=(0.98,0.95,0.61);
rgb(100bp)=(0.98,0.95,0.61)}
\tikzset{_l3gu5dkuv/.code = {\pgfsetadditionalshadetransform{\pgftransformshift{\pgfpoint{0 bp } { 0 bp }  }  \pgftransformrotate{-45 }  \pgftransformscale{2 } }}}
\pgfdeclarehorizontalshading{_416px7xo5} {150bp} {color(0bp)=(transparent!0);
color(37.5bp)=(transparent!0);
color(49.75bp)=(transparent!0);
color(50bp)=(transparent!25);
color(62.5bp)=(transparent!25);
color(100bp)=(transparent!25) }
\pgfdeclarefading{_lt1t64xvl}{\tikz \fill[shading=_416px7xo5,_l3gu5dkuv] (0,0) rectangle (50bp,50bp); }
\tikzset{every picture/.style={line width=0.75pt}} 

\begin{tikzpicture}[x=0.75pt,y=0.75pt,yscale=-1,xscale=1]

\draw  [color={rgb, 255:red, 0; green, 0; blue, 0 }  ,draw opacity=1 ][fill={rgb, 255:red, 255; green, 221; blue, 221 }  ,fill opacity=1 ] (50,150) -- (80,150) -- (80,180) -- (50,180) -- cycle ;
\draw  [fill={rgb, 255:red, 255; green, 221; blue, 221 }  ,fill opacity=1 ] (80,150) -- (110,150) -- (110,180) -- (80,180) -- cycle ;
\draw  [fill={rgb, 255:red, 255; green, 221; blue, 221 }  ,fill opacity=1 ] (110,150) -- (140,150) -- (140,180) -- (110,180) -- cycle ;
\path  [shading=_3k5a8hzom,_pjc13y4it,path fading= _9ypxvc9ks ,fading transform={xshift=2}] (140,150) -- (170,150) -- (170,180) -- (140,180) -- cycle ; 
 \draw  [color={rgb, 255:red, 0; green, 0; blue, 0 }  ,draw opacity=1 ] (140,150) -- (170,150) -- (170,180) -- (140,180) -- cycle ; 

\draw  [fill={rgb, 255:red, 249; green, 243; blue, 156 }  ,fill opacity=0.75 ] (50,182) -- (80,182) -- (80,212) -- (50,212) -- cycle ;
\draw  [fill={rgb, 255:red, 255; green, 221; blue, 221 }  ,fill opacity=1 ] (80,182) -- (110,182) -- (110,212) -- (80,212) -- cycle ;
\draw  [fill={rgb, 255:red, 255; green, 221; blue, 221 }  ,fill opacity=1 ] (110,182) -- (140,182) -- (140,212) -- (110,212) -- cycle ;
\draw  [fill={rgb, 255:red, 255; green, 221; blue, 221 }  ,fill opacity=1 ] (140,182) -- (170,182) -- (170,212) -- (140,212) -- cycle ;
\draw  [fill={rgb, 255:red, 249; green, 243; blue, 156 }  ,fill opacity=0.75 ] (230,182) -- (260,182) -- (260,212) -- (230,212) -- cycle ;
\draw  [fill={rgb, 255:red, 255; green, 221; blue, 221 }  ,fill opacity=1 ] (445.33,100.67) -- (475.33,100.67) -- (475.33,130.67) -- (445.33,130.67) -- cycle ;
\draw  [fill={rgb, 255:red, 172; green, 222; blue, 138 }  ,fill opacity=1 ] (535.33,102.67) -- (565.33,102.67) -- (565.33,132.67) -- (535.33,132.67) -- cycle ;
\draw  [fill={rgb, 255:red, 255; green, 221; blue, 221 }  ,fill opacity=1 ] (475.33,132.67) -- (505.33,132.67) -- (505.33,162.67) -- (475.33,162.67) -- cycle ;
\draw  [fill={rgb, 255:red, 255; green, 221; blue, 221 }  ,fill opacity=1 ] (505.33,132.67) -- (535.33,132.67) -- (535.33,162.67) -- (505.33,162.67) -- cycle ;
\draw  [fill={rgb, 255:red, 172; green, 222; blue, 138 }  ,fill opacity=1 ] (595.33,132.67) -- (625.33,132.67) -- (625.33,162.67) -- (595.33,162.67) -- cycle ;
\draw  [fill={rgb, 255:red, 255; green, 221; blue, 221 }  ,fill opacity=1 ] (475.33,200.67) -- (505.33,200.67) -- (505.33,230.67) -- (475.33,230.67) -- cycle ;
\draw  [fill={rgb, 255:red, 255; green, 221; blue, 221 }  ,fill opacity=1 ] (505.33,200.67) -- (535.33,200.67) -- (535.33,230.67) -- (505.33,230.67) -- cycle ;
\draw  [fill={rgb, 255:red, 172; green, 222; blue, 138 }  ,fill opacity=1 ] (565.33,200.67) -- (595.33,200.67) -- (595.33,230.67) -- (565.33,230.67) -- cycle ;
\draw  [fill={rgb, 255:red, 172; green, 222; blue, 138 }  ,fill opacity=1 ] (445.33,232.67) -- (475.33,232.67) -- (475.33,262.67) -- (445.33,262.67) -- cycle ;
\draw  [fill={rgb, 255:red, 255; green, 221; blue, 221 }  ,fill opacity=1 ] (535.33,232.67) -- (565.33,232.67) -- (565.33,262.67) -- (535.33,262.67) -- cycle ;
\draw  [dash pattern={on 0.84pt off 2.51pt}] (440.33,95.67) -- (658.33,95.67) -- (658.33,165.92) -- (440.33,165.92) -- cycle ;
\draw  [dash pattern={on 0.84pt off 2.51pt}] (441.33,196.67) -- (659.33,196.67) -- (659.33,266.92) -- (441.33,266.92) -- cycle ;
\draw  [fill={rgb, 255:red, 172; green, 222; blue, 138 }  ,fill opacity=1 ] (344.83,181.33) .. controls (344.83,167.53) and (356.03,156.33) .. (369.83,156.33) .. controls (383.64,156.33) and (394.83,167.53) .. (394.83,181.33) .. controls (394.83,195.14) and (383.64,206.33) .. (369.83,206.33) .. controls (356.03,206.33) and (344.83,195.14) .. (344.83,181.33) -- cycle ;
\draw [color={rgb, 255:red, 129; green, 160; blue, 178 }  ,draw opacity=1 ]   (155.29,119.71) -- (370.1,119.82) ;
\draw [color={rgb, 255:red, 129; green, 160; blue, 178 }  ,draw opacity=1 ]   (370.1,119.82) -- (369.83,151.77) ;
\draw [shift={(369.8,154.77)}, rotate = 270.49] [fill={rgb, 255:red, 129; green, 160; blue, 178 }  ,fill opacity=1 ][line width=0.08]  [draw opacity=0] (8.93,-4.29) -- (0,0) -- (8.93,4.29) -- cycle    ;
\draw [color={rgb, 255:red, 129; green, 160; blue, 178 }  ,draw opacity=1 ]   (155.29,119.71) -- (155,149.6) ;
\draw [color={rgb, 255:red, 129; green, 160; blue, 178 }  ,draw opacity=1 ]   (184.79,119.71) -- (184.5,149.6) ;
\draw [color={rgb, 255:red, 129; green, 160; blue, 178 }  ,draw opacity=1 ]   (370.5,240.22) -- (370.05,211.57) ;
\draw [shift={(370,208.57)}, rotate = 89.09] [fill={rgb, 255:red, 129; green, 160; blue, 178 }  ,fill opacity=1 ][line width=0.08]  [draw opacity=0] (8.93,-4.29) -- (0,0) -- (8.93,4.29) -- cycle    ;
\draw [color={rgb, 255:red, 129; green, 160; blue, 178 }  ,draw opacity=1 ]   (184,239.05) -- (370.5,240.22) ;
\draw [color={rgb, 255:red, 129; green, 160; blue, 178 }  ,draw opacity=1 ]   (184.5,212.25) -- (184,239.05) ;
\draw [color={rgb, 255:red, 129; green, 160; blue, 178 }  ,draw opacity=1 ]   (215.5,212.75) -- (215.4,238.4) ;
\draw [color={rgb, 255:red, 122; green, 174; blue, 80 }  ,draw opacity=1 ]   (394.6,173.85) -- (641.6,173.85) ;
\draw [color={rgb, 255:red, 122; green, 174; blue, 80 }  ,draw opacity=1 ]   (610.1,174.35) -- (610.17,167.37) ;
\draw [shift={(610.2,164.37)}, rotate = 90.57] [fill={rgb, 255:red, 122; green, 174; blue, 80 }  ,fill opacity=1 ][line width=0.08]  [draw opacity=0] (8.93,-4.29) -- (0,0) -- (8.93,4.29) -- cycle    ;
\draw [color={rgb, 255:red, 122; green, 174; blue, 80 }  ,draw opacity=1 ]   (550.1,173.35) -- (549.45,136.77) ;
\draw [shift={(549.4,133.77)}, rotate = 88.99] [fill={rgb, 255:red, 122; green, 174; blue, 80 }  ,fill opacity=1 ][line width=0.08]  [draw opacity=0] (8.93,-4.29) -- (0,0) -- (8.93,4.29) -- cycle    ;
\draw [color={rgb, 255:red, 122; green, 174; blue, 80 }  ,draw opacity=1 ]   (579.6,190.35) -- (579.73,196.47) ;
\draw [shift={(579.8,199.47)}, rotate = 268.74] [fill={rgb, 255:red, 122; green, 174; blue, 80 }  ,fill opacity=1 ][line width=0.08]  [draw opacity=0] (8.93,-4.29) -- (0,0) -- (8.93,4.29) -- cycle    ;
\draw  [dash pattern={on 0.84pt off 2.51pt}] (440.67,297.33) -- (658.67,297.33) -- (658.67,367.58) -- (440.67,367.58) -- cycle ;
\draw  [fill={rgb, 255:red, 255; green, 255; blue, 255 }  ,fill opacity=1 ] (564.67,333.33) -- (594.67,333.33) -- (594.67,363.33) -- (564.67,363.33) -- cycle ;
\draw  [fill={rgb, 255:red, 172; green, 222; blue, 138 }  ,fill opacity=1 ] (625.33,132.67) -- (655.33,132.67) -- (655.33,162.67) -- (625.33,162.67) -- cycle ;
\draw  [dash pattern={on 4.5pt off 4.5pt}] (318.33,86.33) -- (668.33,86.33) -- (668.33,375.33) -- (318.33,375.33) -- cycle ;
\draw  [color={rgb, 255:red, 0; green, 0; blue, 0 }  ,draw opacity=1 ][fill={rgb, 255:red, 255; green, 221; blue, 221 }  ,fill opacity=1 ] (48.17,295.5) -- (78.17,295.5) -- (78.17,325.5) -- (48.17,325.5) -- cycle ;
\draw  [fill={rgb, 255:red, 255; green, 221; blue, 221 }  ,fill opacity=1 ] (78.17,295.5) -- (108.17,295.5) -- (108.17,325.5) -- (78.17,325.5) -- cycle ;
\draw  [fill={rgb, 255:red, 255; green, 221; blue, 221 }  ,fill opacity=1 ] (108.17,295.5) -- (138.17,295.5) -- (138.17,325.5) -- (108.17,325.5) -- cycle ;
\draw  [color={rgb, 255:red, 0; green, 0; blue, 0 }  ,draw opacity=1 ][fill={rgb, 255:red, 172; green, 222; blue, 138 }  ,fill opacity=1 ] (138.17,295.5) -- (168.17,295.5) -- (168.17,325.5) -- (138.17,325.5) -- cycle ;
\draw  [fill={rgb, 255:red, 172; green, 222; blue, 138 }  ,fill opacity=1 ] (168.17,295.5) -- (198.17,295.5) -- (198.17,325.5) -- (168.17,325.5) -- cycle ;
\draw  [fill={rgb, 255:red, 172; green, 222; blue, 138 }  ,fill opacity=1 ] (48.17,327.5) -- (78.17,327.5) -- (78.17,357.5) -- (48.17,357.5) -- cycle ;
\draw  [fill={rgb, 255:red, 255; green, 221; blue, 221 }  ,fill opacity=1 ] (78.17,327.5) -- (108.17,327.5) -- (108.17,357.5) -- (78.17,357.5) -- cycle ;
\draw  [fill={rgb, 255:red, 255; green, 221; blue, 221 }  ,fill opacity=1 ] (108.17,327.5) -- (138.17,327.5) -- (138.17,357.5) -- (108.17,357.5) -- cycle ;
\draw  [fill={rgb, 255:red, 255; green, 221; blue, 221 }  ,fill opacity=1 ] (138.17,327.5) -- (168.17,327.5) -- (168.17,357.5) -- (138.17,357.5) -- cycle ;
\draw  [fill={rgb, 255:red, 255; green, 255; blue, 255 }  ,fill opacity=1 ] (168.17,327.5) -- (198.17,327.5) -- (198.17,357.5) -- (168.17,357.5) -- cycle ;
\draw  [fill={rgb, 255:red, 172; green, 222; blue, 138 }  ,fill opacity=1 ] (198.17,327.5) -- (228.17,327.5) -- (228.17,357.5) -- (198.17,357.5) -- cycle ;
\draw  [fill={rgb, 255:red, 172; green, 222; blue, 138 }  ,fill opacity=1 ] (228.17,327.5) -- (258.17,327.5) -- (258.17,357.5) -- (228.17,357.5) -- cycle ;
\draw    (318.38,323.31) -- (266.6,323.34)(318.38,320.31) -- (266.6,320.34) ;
\draw [shift={(258.6,321.85)}, rotate = 359.96] [color={rgb, 255:red, 0; green, 0; blue, 0 }  ][line width=0.75]    (10.93,-3.29) .. controls (6.95,-1.4) and (3.31,-0.3) .. (0,0) .. controls (3.31,0.3) and (6.95,1.4) .. (10.93,3.29)   ;
\draw [shift={(318.38,321.81)}, rotate = 359.96] [color={rgb, 255:red, 0; green, 0; blue, 0 }  ][line width=0.75]    (0,5.59) -- (0,-5.59)   ;
\draw [color={rgb, 255:red, 122; green, 174; blue, 80 }  ,draw opacity=1 ]   (459.6,190.35) -- (459.51,227.75) ;
\draw [shift={(459.5,230.75)}, rotate = 270.14] [fill={rgb, 255:red, 122; green, 174; blue, 80 }  ,fill opacity=1 ][line width=0.08]  [draw opacity=0] (8.93,-4.29) -- (0,0) -- (8.93,4.29) -- cycle    ;
\path  [shading=_7os5ngkt3,_lzgwlebce,path fading= _6g4v1qgqe ,fading transform={xshift=2}] (170,150) -- (200,150) -- (200,180) -- (170,180) -- cycle ; 
 \draw  [color={rgb, 255:red, 0; green, 0; blue, 0 }  ,draw opacity=1 ] (170,150) -- (200,150) -- (200,180) -- (170,180) -- cycle ; 

\path  [shading=_7duo1kxfb,_hodct11gz,path fading= _po9bf0tx0 ,fading transform={xshift=2}] (170,182) -- (200,182) -- (200,212) -- (170,212) -- cycle ; 
 \draw  [color={rgb, 255:red, 0; green, 0; blue, 0 }  ,draw opacity=1 ] (170,182) -- (200,182) -- (200,212) -- (170,212) -- cycle ; 

\path  [shading=_pfd3bzl2l,_uae9mdzdk,path fading= _lt1t64xvl ,fading transform={xshift=2}] (200,182) -- (230,182) -- (230,212) -- (200,212) -- cycle ; 
 \draw  [color={rgb, 255:red, 0; green, 0; blue, 0 }  ,draw opacity=1 ] (200,182) -- (230,182) -- (230,212) -- (200,212) -- cycle ; 

\draw [color={rgb, 255:red, 122; green, 174; blue, 80 }  ,draw opacity=1 ]   (641.6,173.85) -- (641.25,170.55) -- (641.22,167.37) ;
\draw [shift={(641.2,164.37)}, rotate = 89.55] [fill={rgb, 255:red, 122; green, 174; blue, 80 }  ,fill opacity=1 ][line width=0.08]  [draw opacity=0] (8.93,-4.29) -- (0,0) -- (8.93,4.29) -- cycle    ;
\draw [color={rgb, 255:red, 122; green, 174; blue, 80 }  ,draw opacity=1 ]   (394.1,190.35) -- (482.1,190.61) -- (579.6,190.35) ;

\draw (52,153.4) node [anchor=north west][inner sep=0.75pt]    {$c_{1,1}$};
\draw (82,153.4) node [anchor=north west][inner sep=0.75pt]    {$c_{1,2}$};
\draw (112,153.4) node [anchor=north west][inner sep=0.75pt]    {$c_{1,3}$};
\draw (142,153.4) node [anchor=north west][inner sep=0.75pt]    {$c_{1,4}$};
\draw (172,153.4) node [anchor=north west][inner sep=0.75pt]    {$c_{1,5}$};
\draw (52,185.4) node [anchor=north west][inner sep=0.75pt]    {$c_{2,1}$};
\draw (82,185.4) node [anchor=north west][inner sep=0.75pt]    {$c_{2,2}$};
\draw (112,185.4) node [anchor=north west][inner sep=0.75pt]    {$c_{2,3}$};
\draw (142,185.4) node [anchor=north west][inner sep=0.75pt]    {$c_{2,4}$};
\draw (172,185.4) node [anchor=north west][inner sep=0.75pt]    {$c_{2,5}$};
\draw (202,185.4) node [anchor=north west][inner sep=0.75pt]    {$c_{2,6}$};
\draw (232,185.4) node [anchor=north west][inner sep=0.75pt]    {$c_{2,7}$};
\draw (447.33,104.07) node [anchor=north west][inner sep=0.75pt]    {$c_{1,1}$};
\draw (477.33,204.07) node [anchor=north west][inner sep=0.75pt]    {$c_{1,2}$};
\draw (507.33,204.07) node [anchor=north west][inner sep=0.75pt]    {$c_{1,3}$};
\draw (477.33,136.07) node [anchor=north west][inner sep=0.75pt]    {$c_{2,2}$};
\draw (507.33,136.07) node [anchor=north west][inner sep=0.75pt]    {$c_{2,3}$};
\draw (537.33,236.07) node [anchor=north west][inner sep=0.75pt]    {$c_{2,4}$};
\draw (362.83,174.4) node [anchor=north west][inner sep=0.75pt]  [font=\large]  {$\sigma $};
\draw (29.33,151.73) node [anchor=north west][inner sep=0.75pt]  [font=\scriptsize]  {$C_{I_{1}}$};
\draw (28.67,185.07) node [anchor=north west][inner sep=0.75pt]  [font=\scriptsize]  {$C_{I_{2}}$};
\draw (635.33,99.73) node [anchor=north west][inner sep=0.75pt]  [font=\scriptsize]  {$C_{F_{1}}$};
\draw (636.67,201.73) node [anchor=north west][inner sep=0.75pt]  [font=\scriptsize]  {$C_{F_{2}}$};
\draw (356.01,114.77) node  [font=\footnotesize,color={rgb, 255:red, 129; green, 160; blue, 178 }  ,opacity=1 ] [align=left] {\begin{minipage}[lt]{40.33pt}\setlength\topsep{0pt}
reading
\end{minipage}};
\draw (356.68,252.77) node  [font=\footnotesize,color={rgb, 255:red, 129; green, 160; blue, 178 }  ,opacity=1 ] [align=left] {\begin{minipage}[lt]{40.33pt}\setlength\topsep{0pt}
reading
\end{minipage}};
\draw (396.6,176.85) node [anchor=north west][inner sep=0.75pt]  [font=\footnotesize,color={rgb, 255:red, 110; green, 169; blue, 71 }  ,opacity=1 ] [align=left] {writting};
\draw (572.67,347.33) node [anchor=north west][inner sep=0.75pt]  [font=\footnotesize] [align=left] {idle};
\draw (50.17,298.9) node [anchor=north west][inner sep=0.75pt]    {$c_{1,1}$};
\draw (80.17,298.9) node [anchor=north west][inner sep=0.75pt]    {$c_{1,2}$};
\draw (110.17,298.9) node [anchor=north west][inner sep=0.75pt]    {$c_{1,3}$};
\draw (80.17,330.9) node [anchor=north west][inner sep=0.75pt]    {$c_{2,2}$};
\draw (110.17,330.9) node [anchor=north west][inner sep=0.75pt]    {$c_{2,3}$};
\draw (140.17,328.9) node [anchor=north west][inner sep=0.75pt]    {$c_{2,4}$};
\draw (176.17,342.83) node [anchor=north west][inner sep=0.75pt]  [font=\footnotesize] [align=left] {idle};
\draw (200.17,330.9) node [anchor=north west][inner sep=0.75pt]  [font=\footnotesize]  {$c_{3 ,2}$};
\draw (170.17,298.9) node [anchor=north west][inner sep=0.75pt]  [font=\footnotesize]  {$c_{4 ,2}$};
\draw (140.17,298.9) node [anchor=north west][inner sep=0.75pt]  [font=\footnotesize]  {$c_{3 ,1}$};
\draw (323.67,340.83) node [anchor=north west][inner sep=0.75pt]  [font=\footnotesize] [align=left] {conversion\\procedure};
\draw (269.95,305.14) node [anchor=north west][inner sep=0.75pt]  [font=\scriptsize] [align=left] {physically};
\draw (50.17,330.9) node [anchor=north west][inner sep=0.75pt]  [font=\footnotesize]  {$c_{4 ,1}$};
\draw (230.17,330.9) node [anchor=north west][inner sep=0.75pt]  [font=\footnotesize]  {$c_{3 ,3}$};
\draw (537.33,106.07) node [anchor=north west][inner sep=0.75pt]  [font=\footnotesize]  {$c_{3 ,1}$};
\draw (567.33,204.07) node [anchor=north west][inner sep=0.75pt]  [font=\footnotesize]  {$c_{4 ,2}$};
\draw (447.33,236.07) node [anchor=north west][inner sep=0.75pt]  [font=\footnotesize]  {$c_{4 ,1}$};
\draw (597.33,136.07) node [anchor=north west][inner sep=0.75pt]  [font=\footnotesize]  {$c_{3 ,2}$};
\draw (627.33,136.07) node [anchor=north west][inner sep=0.75pt]  [font=\footnotesize]  {$c_{3 ,3}$};
\draw (395.1,161.1) node [anchor=north west][inner sep=0.75pt]  [font=\scriptsize]  {$\sigma _{1}$};
\draw (396.1,193.75) node [anchor=north west][inner sep=0.75pt]  [font=\scriptsize]  {$\sigma _{2}$};
\end{tikzpicture}
\caption{A $(2,2)_q$ convertible code for Example 1. Each block is a physically independent device and stores a single symbol. The red devices store unchanged symbols. The blue devices store read symbols. The green devices store written symbols. Notice that the symbols in yellow devices are overwritten by new symbols or released after the conversion procedure, which are called retired symbols in \cite{Maturana2020a}. Moreover, there is an idle device, that is released symbol after conversion procedure, which is exactly the storage overhead saved.}
\label{fig_conversion}
\end{figure}
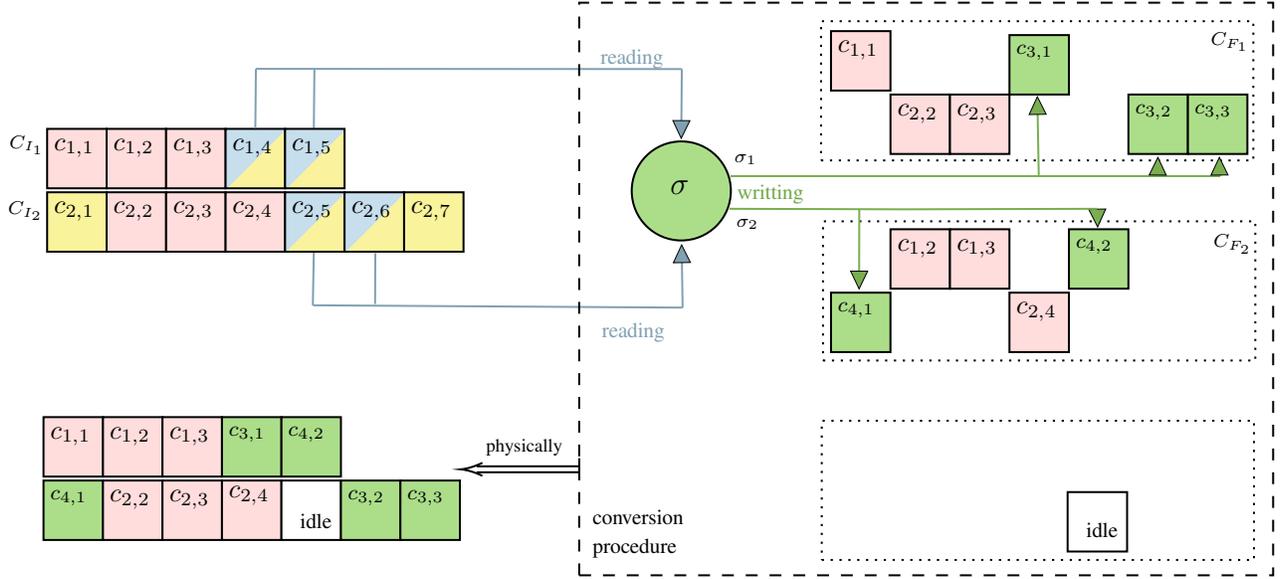

In the storage systems, keeping unchanged symbols has many practical benefits because those symbols can remain in the same physical location (e.g. the red blocks in Figure \ref{fig_conversion})  in the conversion procedure. Moreover, the number of unchanged symbols and the number of written symbols are closely related, i.e., $|\written_{j}|=n_{F_j}-\sum_{i\in[t_1]}|\unchanged_{i,j}|$ according to \eqref{eqn_5}.  For generalized convertible codes, we also introduce the following definition to capture codes that maximize the number of unchanged symbols, which is termed stability in \cite{Maturana2020a}.

\begin{definition}[Stable Generalized Convertible Code]\label{def: stable}
A $(t_1,t_2)_q$ generalized convertible code $\bm{\mathcal{C}}$ is said to be stable if it has the maximum number of unchanged symbols overall $(t_1,t_2)_q$ generalized convertible codes.
\end{definition}

During the conversion procedure, the read symbols and written symbols are involved in I/O overhead physical systems.  In order to measure of I/O performance of the generalized convertible codes, we also define the access cost, which is first introduced in \cite{Maturana2020a}.

\begin{definition}[Access Cost]\label{def_access_cost}
For a $(t_1,t_2)_q$ generalized convertible code $\bm{\mathcal{C}}$, the access cost  $\rho$ is defined as the sum of read access cost $\rho_r=\sum_{i\in[t_1]}|\bigcup_{j\in[t_2]}\mathcal{R}_{i,j}|$ and write access cost $\rho_w=\sum_{j\in[t_2]}(n_{F_j}-\sum_{i\in[t_1]}|\unchanged_{i,j}|)$, where the read access cost is the number of symbols in initial codes used in the conversion procedure functions and the write access cost counts the number of symbols written in the final codes.
\end{definition}

Clearly, there exist $k_{I_i}$ symbols that can recover the entire initial code $\mathcal{C}_{I_i}$ for $i\in[t_1]$ since $\mathcal{C}_{I_i}$ is $k_{I_i}$ dimensional linear code. Therefore, we can always read at most $k_{I_i}$ symbols of initial code $\mathcal{C}_{I_i}$ during the conversion procedure. In this trivial case, the conversion method for initial code $\mathcal{C}_{I_i}$ is called \textit{default approach}, i.e., $|\bigcup_{j\in[t_2]}\mathcal{R}_{i,j}|=k_{I_i}$.

We give a $(2,2)_q$ generalized convertible code below to  illustrate the above conversion process,

\begin{example}\label{Exam-1}
Assume that  the initial codes are $[5,3]_q$ code $\mathcal{C}_{I_1}$  and $[7,4]_q$ code $\mathcal{C}_{I_2}$, and  the final codes are $[6,4]_q$ code $\mathcal{C}_{F_1}$ and  $[5,3]$ code $\mathcal{C}_{F_2}$. For any two initial codewords $C_{I_1}=(c_{1,1},c_{1,2},\cdots,c_{1,5})\in\mathcal{C}_{I_1}$ and $C_{I_2}=(c_{2,1},c_{2,2},\cdots,c_{2,7})\in\mathcal{C}_{I_2}$, the sets of unchanged, read and written symbols are
\begin{itemize}
\item $\unchanged_{1,1}=\{(1,1)\},\unchanged_{1,2}=\{(1,2),(1,3)\},\unchanged_{2,1}=\{(2,2),(2,3)\},\unchanged_{2,2}=\{(2,4)\}$;
\item $\reading_{1,1}=\{(1,4),(1,5)\},\reading_{1,2}=\{(1,5)\},\reading_{2,1}=\{(2,5),(2,6)\},\reading_{2,2}=\{(2,5)\}$;
\item $\written_{1}=\{(3,1),(3,2),(3,3)\}, \written_{2}=\{(4,1),(4,2)\}$.
\end{itemize}
Then, two final codewords  $C_{F_1}=(c_{1,1},c_{2,2},c_{2,3},c_{3,1},c_{3,2},c_{3,3})\in\mathcal{C}_{F_1}$ and $C_{F_2}=(c_{4,1},c_{1,2},c_{1,3},c_{2,4},c_{4,2})\in\mathcal{C}_{F_2}$ are obtained by the conversion procedure function $\sigma=(\sigma_1,\sigma_2)$ is shown in Figure \ref{fig_conversion} and
\begin{align*}
\sigma_1((c_{1,4},c_{1,5},c_{2,5},c_{2,6})) &= (c_{1,4},c_{1,5}+c_{2,5},c_{1,4}+c_{2,6})=(c_{3,1},c_{3,2},c_{3,3}),\\
\sigma_2((c_{1,5},c_{2,5})) &= (c_{1,5}+c_{2,5},c_{1,5}) = (c_{4,1},c_{4,2}).
\end{align*}
Clearly, the read access cost is $\rho_r=4$ and the write access cost is $\rho_w=5$.
\end{example}

By the conversion procedure, minimizing the number of written symbols is equivalent to maximizing the number of unchanged symbols. And motivation is to minimize the access cost for MDS convertible codes. Therefore, we introduce the following upper bound on the number of unchanged symbols.

\begin{lemma}\label{prop_1}
Let  $\bm{\mathcal{C}}$ be a $(t_1,1)_q$  MDS generalized  merge-convertible code then $|\unchanged_{i,1}|\le k_{I_i}$ for $1\le i\le t_1$. Let $\bm{\mathcal{C}}$ be a $(1,t_2)_q$  MDS generalized split-convertible code then $|\unchanged_{1,j}|\le k_{F_j}$ for $1\le j\le t_2$ .
    \end{lemma}
\begin{IEEEproof}
Assume $|\unchanged_{i,1}|>k_{I_i}$ for generalized $(t_1,1)_q$ merge-convertible code $\bm{\mathcal{C}}$, where $\sum_{i=1}^{t_1} k_{I_i}=k_{F_1}$.
Let $G_i=(G_{i,1},G_{i,2},$ $\cdots,G_{i,n_{I_i}})$ be a generator matrix of $\mathcal{C}_{I_i}$.
Note that  the  code symbols $\mathcal{C}_{I_i}|_{\unchanged_{i,1}}$ are corresponding to $\{G_{\ell}\}_{\ell\in\unchanged_{i,1}}$. Thus, these  $|\unchanged_{i,1}|$ column vectors
 are linear dependent over $\mathbb{F}_q$. But,  they are also in a generator matrix of code $\mathcal{C}_{F_1}$ according to \eqref{eqn_1}, which contradicts that $\mathcal{C}_{F_1}$ is an MDS code with dimension $k_{F_1}\ge|\unchanged_{i,1}|> k_{I_i}$.  The proof for a generalized split-convertible code is similar, so we omit it here.
\end{IEEEproof}

\section{Lower Bound on Access Cost}\label{sec_bounds}
In this section, we establish a new lower bound on access cost
of  MDS generalized merge-convertible codes and  MDS generalized split-convertible codes, which generalizes the known results in \cite{Maturana2020} and \cite{Maturana2022}, respectively.

We begin with the following property of MDS codes used in this section.

\begin{lemma}\label{lem:fact}
    Let $\mathcal{C}$ and $\mathcal{C}'$ be $[n,k]_q$ and $[n',k']_q$ codes, respectively. Assume that $\mathcal{C}|_\mathcal{A}$  can be linearly generated by $\mathcal{C}'|_\mathcal{B}$ for $\mathcal{A}\subseteq[n]$ and $\mathcal{B}\subseteq[n']$.
    \begin{enumerate}
     \item[1)] If $\mathcal{C}$ is an MDS code and $\mathcal{C}\ne\mathcal{C}'$,  $|\mathcal{A}|>\min\{k,|\mathcal{B}|\}$, then $|\mathcal{B}|\ge k$;
     \item[2)] When $\mathcal{C}$ is an MDS code and $\mathcal{C}=\mathcal{C}'$, $\mathcal{A}\not\subseteq\mathcal{B}$,  assumption holds if and only if $|\mathcal{B}|\ge k$.
    \end{enumerate}
\end{lemma}
\begin{IEEEproof}
1) We are done if $|\mathcal{B}|\ge k$. Assume on the contrary that $|\mathcal{B}|<k$, i.e., $|\mathcal{A}|>\min\{k,|\mathcal{B}|\}= |\mathcal{B}|$.

According to \eqref{eqn_generate}, there exists a $k\times k'$ matrix $U$ and a $|\mathcal{B}|\times |\mathcal{A}|$ matrix $V$ over $\mathbb{F}_q$ such that $G|_{\mathcal{A}}=UG'|_{\mathcal{B}}V$, where $G$ and $G'$ are generator matrices of $\mathcal{C}$ and $\mathcal{C}'$, respectively. On one hand, we have
\begin{align*}
    \rank(G|_{\mathcal{A}})
    &\le\min\{\rank(U),\rank(G'|_{\mathcal{B}}),\rank(V)\}\\
    &\le\min\{k,k',|\mathcal{A}|,|\mathcal{B}|\}\\
    &=\min\{k',|\mathcal{B}|\}.
\end{align*}
On the other hand, we can obtain  $\rank(G|_{\mathcal{A}})=\min\{k,|\mathcal{A}|\}$ since $\mathcal{C}$ is an MDS code with dimension $k$. Thus,
\begin{eqnarray*}
\min\{k,|\mathcal{A}|\}\le \min\{k',|\mathcal{B}|\},
\end{eqnarray*}
which implies either $k\le |\mathcal{B}|$ or $|\mathcal{A}|\le |\mathcal{B}|$, a contradiction.

2) This is directly from the property of an MDS code $\mathcal{C}$, that any symbol from $\mathcal{C}$ can be recovered by other $t$ symbols if and only if $t\ge k$.
\end{IEEEproof}

\subsection{Merge Regime}

In the merge regime, the final code defined by \eqref{eqn_5} turns into
\begin{equation*}
\mathcal{C}_{F_1}
=\left\{ (\bm{x}_1,\sigma_1(\bm{x}_2)) : \bm{x}\in\bigotimes_{i\in[t_1]}\mathcal{C}_{I_i}, \bm{x}_1=\bm{x}|_{\bigcup_{i\in[t_1]}\unchanged_{i,1}}, \bm{x}_2=\bm{x}|_{\bigcup_{i\in[t_1]}\reading_{i,1}} \right\}
\end{equation*}
with dimension $\sum_{i\in[t_1]}k_{I_i}=k_{F_1}$ and the cardinality of the written symbols is
\begin{equation}\label{eqn_write_cost}
|\written_1|=n_{F_1}-\sum_{i\in[t_1]}|\unchanged_{i,1}|.
\end{equation}

We have the following lower bound on access cost of MDS generalized merge-convertible codes.

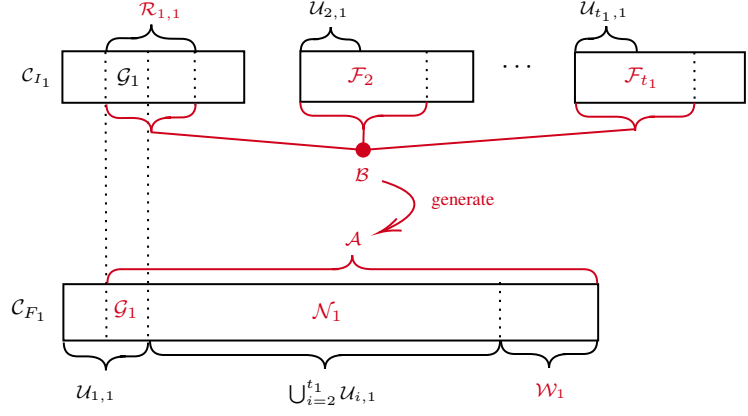
\begin{figure}
\centering
\tikzset{every picture/.style={line width=0.75pt}} 
\begin{tikzpicture}[x=0.75pt,y=0.75pt,yscale=-1,xscale=1]
\draw   (40.5,170.5) -- (310.2,170.5) -- (310.2,198.95) -- (40.5,198.95) -- cycle ;
\draw   (40,52.95) -- (131.7,52.95) -- (131.7,78.95) -- (40,78.95) -- cycle ;
\draw   (160,52.95) -- (247.2,52.95) -- (247.2,78.45) -- (160,78.45) -- cycle ;
\draw   (298.5,53.45) -- (384.2,53.45) -- (384.2,78.95) -- (298.5,78.95) -- cycle ;
\draw  [dash pattern={on 0.84pt off 2.51pt}]  (260.7,169.45) -- (260.7,198.45) ;
\draw  [dash pattern={on 0.84pt off 2.51pt}]  (83.2,52.75) -- (83.2,198.45) ;
\draw   (40.7,199.95) .. controls (40.7,204.62) and (43.03,206.95) .. (47.7,206.95) -- (51.95,206.95) .. controls (58.62,206.95) and (61.95,209.28) .. (61.95,213.95) .. controls (61.95,209.28) and (65.28,206.95) .. (71.95,206.95)(68.95,206.95) -- (76.2,206.95) .. controls (80.87,206.95) and (83.2,204.62) .. (83.2,199.95) ;
\draw  [dash pattern={on 0.84pt off 2.51pt}]  (61.7,53.25) -- (62.2,197.95) ;
\draw   (107.2,53.75) .. controls (107.25,49.08) and (104.95,46.72) .. (100.28,46.67) -- (94.78,46.61) .. controls (88.11,46.54) and (84.81,44.17) .. (84.86,39.5) .. controls (84.81,44.17) and (81.45,46.46) .. (74.78,46.39)(77.78,46.42) -- (69.28,46.33) .. controls (64.61,46.28) and (62.25,48.58) .. (62.2,53.25) ;
\draw  [dash pattern={on 0.84pt off 2.51pt}]  (223.7,52.95) -- (223.85,78.2) ;
\draw  [dash pattern={on 0.84pt off 2.51pt}]  (358.7,53.45) -- (358.85,78.7) ;
\draw  [dash pattern={on 0.84pt off 2.51pt}]  (106.7,52.95) -- (106.85,80.2) ;
\draw  [color={rgb, 255:red, 208; green, 2; blue, 27 }  ,draw opacity=1 ] (62.2,79.75) .. controls (62.25,84.42) and (64.61,86.72) .. (69.28,86.67) -- (74.78,86.61) .. controls (81.45,86.54) and (84.81,88.83) .. (84.86,93.5) .. controls (84.81,88.83) and (88.11,86.46) .. (94.78,86.39)(91.78,86.42) -- (100.28,86.33) .. controls (104.95,86.28) and (107.25,83.92) .. (107.2,79.25) ;
\draw  [color={rgb, 255:red, 208; green, 2; blue, 27 }  ,draw opacity=1 ] (160,78.5) .. controls (160,83.17) and (162.33,85.5) .. (167,85.5) -- (181.85,85.48) .. controls (188.52,85.48) and (191.85,87.81) .. (191.86,92.48) .. controls (191.85,87.81) and (195.18,85.48) .. (201.85,85.47)(198.85,85.47) -- (216.7,85.46) .. controls (221.37,85.46) and (223.7,83.13) .. (223.7,78.46) ;
\draw  [color={rgb, 255:red, 208; green, 2; blue, 27 }  ,draw opacity=1 ] (299,79) .. controls (299.01,83.67) and (301.34,86) .. (306.01,85.99) -- (318.86,85.98) .. controls (325.53,85.97) and (328.86,88.3) .. (328.86,92.97) .. controls (328.86,88.3) and (332.19,85.97) .. (338.86,85.97)(335.86,85.97) -- (351.71,85.95) .. controls (356.38,85.95) and (358.71,83.62) .. (358.7,78.95) ;
\draw  [color={rgb, 255:red, 208; green, 2; blue, 27 }  ,draw opacity=1 ] (309.7,169.95) .. controls (309.7,165.28) and (307.37,162.95) .. (302.7,162.95) -- (196.2,162.95) .. controls (189.53,162.95) and (186.2,160.62) .. (186.2,155.95) .. controls (186.2,160.62) and (182.87,162.95) .. (176.2,162.95)(179.2,162.95) -- (69.7,162.95) .. controls (65.03,162.95) and (62.7,165.28) .. (62.7,169.95) ;
\draw [color={rgb, 255:red, 208; green, 2; blue, 27 }  ,draw opacity=1 ]   (84.5,93.8) -- (191.7,102.75) ;
\draw [color={rgb, 255:red, 208; green, 2; blue, 27 }  ,draw opacity=1 ]   (192,91) -- (191.7,102.75) ;
\draw [shift={(191.7,102.75)}, rotate = 91.46] [color={rgb, 255:red, 208; green, 2; blue, 27 }  ,draw opacity=1 ][fill={rgb, 255:red, 208; green, 2; blue, 27 }  ,fill opacity=1 ][line width=0.75]      (0, 0) circle [x radius= 3.35, y radius= 3.35]   ;
\draw [color={rgb, 255:red, 208; green, 2; blue, 27 }  ,draw opacity=1 ]   (329.2,92.25) -- (191.7,102.75) ;
\draw [color={rgb, 255:red, 208; green, 2; blue, 27 }  ,draw opacity=1 ]   (201.2,118.45) .. controls (226.42,126.21) and (219.18,136.32) .. (201.38,143.77) ;
\draw [shift={(199.7,144.45)}, rotate = 338.46] [color={rgb, 255:red, 208; green, 2; blue, 27 }  ,draw opacity=1 ][line width=0.75]    (10.93,-3.29) .. controls (6.95,-1.4) and (3.31,-0.3) .. (0,0) .. controls (3.31,0.3) and (6.95,1.4) .. (10.93,3.29)   ;
\draw   (84.2,199.25) .. controls (84.2,203.92) and (86.53,206.25) .. (91.2,206.25) -- (162.45,206.25) .. controls (169.12,206.25) and (172.45,208.58) .. (172.45,213.25) .. controls (172.45,208.58) and (175.78,206.25) .. (182.45,206.25)(179.45,206.25) -- (253.7,206.25) .. controls (258.37,206.25) and (260.7,203.92) .. (260.7,199.25) ;
\draw   (261.2,198.75) .. controls (261.2,203.42) and (263.53,205.75) .. (268.2,205.75) -- (275.45,205.75) .. controls (282.12,205.75) and (285.45,208.08) .. (285.45,212.75) .. controls (285.45,208.08) and (288.78,205.75) .. (295.45,205.75)(292.45,205.75) -- (302.7,205.75) .. controls (307.37,205.75) and (309.7,203.42) .. (309.7,198.75) ;
\draw   (190.2,53.25) .. controls (190.24,49.13) and (188.2,47.05) .. (184.08,47.01) -- (184.08,47.01) .. controls (178.2,46.95) and (175.28,44.86) .. (175.32,40.75) .. controls (175.28,44.86) and (172.32,46.89) .. (166.44,46.84)(169.08,46.86) -- (166.44,46.84) .. controls (162.32,46.8) and (160.24,48.84) .. (160.2,52.95) ;
\draw   (329.7,53.25) .. controls (329.74,49) and (327.63,46.85) .. (323.38,46.81) -- (323.38,46.81) .. controls (317.3,46.75) and (314.28,44.59) .. (314.32,40.34) .. controls (314.28,44.59) and (311.22,46.69) .. (305.14,46.63)(307.88,46.66) -- (305.14,46.63) .. controls (300.89,46.59) and (298.74,48.7) .. (298.7,52.95) ;
\draw (277,218.9) node [anchor=north west][inner sep=0.75pt]  [font=\scriptsize,color={rgb, 255:red, 208; green, 2; blue, 27 }  ,opacity=1 ]  {$\mathcal{W}_{1}$};
\draw (46,219.9) node [anchor=north west][inner sep=0.75pt]  [font=\scriptsize]  {$\mathcal{U}_{1,1}$};
\draw (152.5,217.4) node [anchor=north west][inner sep=0.75pt]  [font=\scriptsize]  {$\bigcup _{i=2}^{t_{1}}\mathcal{U}_{i,1}$};
\draw (77,26.9) node [anchor=north west][inner sep=0.75pt]  [font=\scriptsize,color={rgb, 255:red, 208; green, 2; blue, 27 }  ,opacity=1 ]  {$\mathcal{R}_{1,1}$};
\draw (163.5,26.4) node [anchor=north west][inner sep=0.75pt]  [font=\scriptsize]  {$\mathcal{U}_{2,1}$};
\draw (300,26.35) node [anchor=north west][inner sep=0.75pt]  [font=\scriptsize]  {$\mathcal{U}_{t_{1} ,1}$};
\draw (260,56.4) node [anchor=north west][inner sep=0.75pt]    {$\cdots $};
\draw (242.6,128.97) node  [font=\scriptsize] [align=left] {\begin{minipage}[lt]{24.62pt}\setlength\topsep{0pt}
\textcolor[rgb]{0.82,0.01,0.11}{generate}
\end{minipage}};
\draw (186,110.4) node [anchor=north west][inner sep=0.75pt]  [font=\scriptsize,color={rgb, 255:red, 208; green, 2; blue, 27 }  ,opacity=1 ]  {$\mathcal{\textcolor[rgb]{0.82,0.01,0.11}{B}}$};
\draw (181,142.9) node [anchor=north west][inner sep=0.75pt]  [font=\scriptsize]  {$\mathcal{\textcolor[rgb]{0.82,0.01,0.11}{A}}$};
\draw (13,176.9) node [anchor=north west][inner sep=0.75pt]  [font=\footnotesize]  {$\mathcal{C}_{F_{1}}$};
\draw (18,59.9) node [anchor=north west][inner sep=0.75pt]  [font=\scriptsize]  {$\mathcal{C}_{I_{1}}$};
\draw (66,60.4) node [anchor=north west][inner sep=0.75pt]  [font=\footnotesize]  {$\mathcal{G}_{1}$};
\draw (64.5,177.4) node [anchor=north west][inner sep=0.75pt]  [font=\footnotesize,color={rgb, 255:red, 208; green, 2; blue, 27 }  ,opacity=1 ]  {$\mathcal{G}_{1}$};
\draw (165,178.9) node [anchor=north west][inner sep=0.75pt]  [font=\footnotesize,color={rgb, 255:red, 208; green, 2; blue, 27 }  ,opacity=1 ]  {$\mathcal{N}_{1}$};
\draw (182.5,59.9) node [anchor=north west][inner sep=0.75pt]  [font=\footnotesize,color={rgb, 255:red, 208; green, 2; blue, 27 }  ,opacity=1 ]  {$\mathcal{F}_{2}$};
\draw (322,59.9) node [anchor=north west][inner sep=0.75pt]  [font=\footnotesize,color={rgb, 255:red, 208; green, 2; blue, 27 }  ,opacity=1 ]  {$\mathcal{F}_{t_{1}}$};
\end{tikzpicture}
\caption{The relationship among code symbols used in the proof of Theorem \ref{thm: merge-convertible: lower bound} for $i=1$, where $\mathcal{F}_j\subseteq\{j\}\times[n_{I_j}]$ such that $|\mathcal{F}_j|=k_{I_j}, \unchanged_{j,1}\subseteq\mathcal{F}_j$ for $j\in[t_1]\setminus\{1\}$, and $\mathcal{G}_1=\unchanged_{1,1}\cap\reading_{1,1}$, and $\mathcal{N}_1=\bigcup_{j\in[t_1]\setminus\{1\}}\unchanged_{j,1}$.}
\label{fig:enter-label1}
\end{figure}

\begin{theorem}\label{thm: merge-convertible: lower bound}
    Let $\bm{\mathcal{C}}$ be a $(t_1,1)_q$  MDS generalized merge-convertible code, then
    \begin{equation}\label{Eqn_R_Bound}
    |\reading_{i,1}|\ge
    \begin{cases}
     r_{F_1},&\quad \text{if }r_{F_1} \le \min\{k_{I_i},r_{I_i}\},\\
     k_{I_i},&\quad \text{otherwise}
    \end{cases}
    \end{equation}
    for all $i\in [t_1]$. That is,
    the access cost $\rho$ satisfies
    \begin{equation}\label{Eqn_MC_ac_Bound}
        \rho \ge r_{F_1}+\sum_{\substack{i\in [t_1],\,r_{F_1} \le\\ \min\{ k_{I_i},r_{I_i} \}}} r_{F_1} + \sum_{\substack{i\in [t_1],\,r_{F_1} >\\ \min\{ k_{I_i},r_{I_i} \}}} k_{I_i},
    \end{equation}
where the equality is achieved only if the equality in \eqref{Eqn_R_Bound} is attained and
$|\unchanged_{i,1}|=k_{I_i}$ for all $i\in [t_1]$.
\end{theorem}
\begin{IEEEproof}
Firstly, we prove that for given $|\unchanged_{i,1}|$,   $|\reading_{i,1}|$ ($i\in [t_1]$), satisfies
    \begin{equation}\label{Eqn_MC_bound}
    |\reading_{i,1}|\ge
    \begin{cases}
     k_{I_i}  + r_{F_1}- |\unchanged_{i,1}|,&\quad \text{if }r_{F_1} \le \min\{|\unchanged_{i,1}|,r_{I_i}\},\\
     k_{I_i},&\quad \text{otherwise.}
    \end{cases}
    \end{equation}

Before proving it, we introduce some sets of code symbols for $(t_1,1)_q$  MDS generalized merge-convertible code $\bm{\mathcal{C}}$.  By Lemma \ref{prop_1},  we can always choose a set  $\mathcal{F}_i\subseteq\{i\}\times[n_{I_i}]$ for each $i\in [t_1]$ such that $|\mathcal{F}_i|=k_{I_i}$ and $\unchanged_{i,1}\subseteq\mathcal{F}_i$. For $i\in[t_1]$, define
\begin{align*}
\mathcal{G}_i&\triangleq\unchanged_{i,1}\cap\reading_{i,1}\subseteq\reading_{i,1},\\
\mathcal{N}_i&\triangleq\bigcup_{j\in[t_1]\setminus\{i\}}\unchanged_{j,1} \subseteq\bigcup_{j\in[t_1]\setminus\{i\}}\mathcal{F}_j.
\end{align*}
Figure \ref{fig:enter-label1} intuitively shows the relationship among the above code symbols.

Recall that $\mathcal{C}_{F_1}|_{\written_1}$ is generated from $\bigotimes_{i\in[t_1]}\mathcal{C}_{I_i}|_{\reading_{i,1}}$ in the conversion procedure. Whereas, $\mathcal{C}_{I_i}|_{\reading_{i,1}}$ can be generated from $\mathcal{C}_{I_i}|_{\mathcal{F}_i}$ for every $i\in[t_1]$
since $\mathcal{C}_{I_i}$ is an MDS code. Then, $\mathcal{C}_{F_1}|_{\written_1}$ can be generated by $\bigotimes_{j\in[t_1]\setminus\{i\}}\mathcal{C}_{I_j}|_{\mathcal{F}_j}\otimes\mathcal{C}_{I_i}|_{\reading_{i,1}}$. Herein, we apply a fact that the generator matrix of $\mathcal{C}_1\bigotimes \mathcal{C}_2$ is $G=\left(\begin{array}{cc}G_1 & O \\ O & G_2\end{array}\right)$, where $G_1$ and $G_2$ are generator matrices of $\mathcal{C}_1$ and $\mathcal{C}_2$, respectively.

In
addition,  $\mathcal{C}_{F_1}|_{\mathcal{G}_i\cup\mathcal{N}_i}$ can be generated by $\bigotimes_{j\in[t_1]\setminus\{i\}}\mathcal{C}_{I_j}|_{\mathcal{F}_j}\otimes\mathcal{C}_{I_i}|_{\reading_{i,1}}$ as well, by \eqref{eqn_1} and the fact that $\mathcal{G}_i,\mathcal{N}_i$ are coordinates sets of unchanged symbols, i.e.,  $\mathcal{C}_{F_1}|_{\mathcal{G}_i\cup\mathcal{N}_i}= \bigotimes_{j\in[t_1]\setminus\{i\}}\mathcal{C}_{I_j}|_{\unchanged_{j,1}}\otimes\mathcal{C}_{I_i}|_{\mathcal{G}_i}$. This is to say that $\mathcal{C}_{F_1}|_{\written_1\cup\mathcal{G}_i\cup\mathcal{N}_i}$ can be generated by $\bigotimes_{j\in[t_1]\setminus\{i\}}\mathcal{C}_{I_j}|_{\mathcal{F}_j}\otimes\mathcal{C}_{I_i}|_{\reading_{i,1}}$ for every $i\in[t_1]$. Note that
\begin{align*}
|\mathcal{A}|
&\triangleq|\written_1\cup\mathcal{G}_i\cup\mathcal{N}_i|\\
&=|\written_1|+|\unchanged_{i,1}\cap\reading_{i,1}|+\left|\bigcup_{j\in[t_1]\setminus\{i\}}\unchanged_{j,1}\right|\\
&=n_{F_1}-\sum_{i\in[t_1]}|\unchanged_{i,1}|+|\unchanged_{i,1}\cap\reading_{i,1}|+\sum_{j\in[t_1]\setminus\{i\}}|\unchanged_{j,1}|\\
&= k_{F_1}+r_{F_1}-|\unchanged_{i,1}|+|\unchanged_{i,1}\cap\reading_{i,1}|
\end{align*}
and
\begin{equation*}
|\mathcal{B}|
\triangleq\sum_{j\in[t_2]\setminus\{i\}}|\mathcal{F}_j|+|\reading_{i,1}|
=\sum_{j\in[t_2]\setminus\{i\}}k_{I_j}+|\reading_{i,1}|.
\end{equation*}

Now, according to Lemma \ref{lem:fact}-1, to prove $|\reading_{i,1}|\ge k_{I_i}$ we only need to prove
$|A|>\min\{k_F,|\mathcal{B}|\}$.
And we prove \eqref{Eqn_MC_bound} in the following three cases.

\textbf{Case 1:} $r_{F_1}>|\unchanged_{i,1}|$ for $i\in[t_1]$. Clearly, $|\mathcal{A}|>k_{F_1}$, then $|\reading_{i,1}|\ge k_{I_i}$.

\textbf{Case 2:} $r_{F_1}>r_{I_i}$ for $i\in[t_1]$. Then,
    \begin{align*}
        |\mathcal{A}|>~&k_{F_1}+r_{I_i}-|\unchanged_{i,1}|+|\reading_{i,1}\cap\unchanged_{i,1}|\\
        =~&\sum_{j\in[t_1]\setminus\{i\}}k_{I_j}+n_{I_i}-|\unchanged_{i,1}|+|\reading_{i,1}\cap\unchanged_{i,1}|\\
        =~&\sum_{j\in[t_1]\setminus\{i\}}k_{I_j}+|(\{i\}\times[n_{I_i}])\setminus\unchanged_{i,1}|+|\reading_{i,1}\cap\unchanged_{i,1}|\\
        \ge~&\sum_{j\in[t_1]\setminus\{i\}}k_{I_j}+|\reading_{i,1}\cap((\{i\}\times[n_{I_i}])\setminus\unchanged_{i,1})|+|\reading_{i,1}\cap\unchanged_{i,1}|\\
        =~&\sum_{j\in[t_1]\setminus\{i\}}k_{I_j}+|\reading_{i,1}|\\
        =~&|\mathcal{B}|.
    \end{align*}
  Therefore, $|\reading_{i,1}|\ge k_{I_i}$.

   \textbf{Case 3:} $r_{F_1} \le \min\{|\unchanged_{i,1}|,r_{I_i}\}$ for $i\in[t_1]$.     Assume on the contrary that the read access cost $|\reading_{i,1}|<k_{I_i}+ r_{F_1}-|\unchanged_{i,1}| $. Then, we have
   \begin{align*}
       |\mathcal{B}|
       <k_{F_1}+r_{F_1}-|\unchanged_{i,1}|\le |\mathcal{A}|.
   \end{align*}
Thus, $|\reading_{i,1}|\ge k_{I_i}$. But the assumptions $r_{F_1}-|\unchanged_{i,1}|\le0$ and $|\reading_{i,1}|<k_{I_i}+ r_{F_1}-|\unchanged_{i,1}| $ imply that $k_{I_i}>|\reading_{i,1}|\ge k_{I_i}$ which is impossible. Hence, $|\reading_{i,1}|\ge k_{I_i}+ r_{F_1}-|\unchanged_{i,1}| $.

Combining the above three cases, we obtain the inequality \eqref{Eqn_MC_bound}. According to Lemma \ref{prop_1}, $|\unchanged_{i,1}|\le k_{I_i}$ for $i\in[t_1]$. Thus, we get
\eqref{Eqn_R_Bound} and \eqref{Eqn_MC_ac_Bound} by maximizing each $|\unchanged_{i,1}|$ in \eqref{Eqn_MC_bound} and \eqref{eqn_write_cost}, i.e., setting $|\unchanged_{i,1}|= k_{I_i}$ for $i\in[t_1]$. This is to say that the lower bound given by \eqref{Eqn_R_Bound} and \eqref{Eqn_MC_ac_Bound} can be achieved with equality only if $|\unchanged_{i,1}|= k_{I_i}$ for $i\in[t_1]$.
\end{IEEEproof}

\begin{definition}\label{Def_access_optimal}
A $(t_1,1)_q$  MDS generalized merge-convertible code $\bm{\mathcal{C}}$  is said to be access-optimal if the read access cost and access cost respectively achieve the lower bounds with equality in \eqref{Eqn_R_Bound} and \eqref{Eqn_MC_ac_Bound}.
\end{definition}

The following corollary is obvious from Theorem \ref{thm: merge-convertible: lower bound}
and Definition \ref{def: stable}.

\begin{corollary}\label{cor_stable}
All $(t_1,1)_q$ access-optimal MDS generalized merge-convertible codes are stable.
\end{corollary}

Theorem \ref{thm: merge-convertible: lower bound}  covers the known bound \cite[Theorem 9]{Maturana2022} by setting $\mathcal{C}_{F_1}=\mathcal{C}_{F_2}=\cdots=\mathcal{C}_{F_{t_2}}$.

\begin{corollary}[\cite{Maturana2022}, Theorem 9]\label{cor: merge-convertible 1}
    For all $(t_1,1)_q$ MDS generalized merge-convertible codes such that $\mathcal{C}_{I_1}=\mathcal{C}_{I_2}=\cdots=\mathcal{C}_{I_{t_1}}$, the access cost $\rho$ satisfies
    \begin{equation*}
        \rho \ge \begin{cases}
                   t_1 r_{F_1}+r_{F_1}, & \mbox{if } r_{F_1} \le \min\{k_{I_1},r_{I_1}\}, \\
                   t_1 k_{I_1}+r_{F_1}, & \mbox{otherwise}.
                 \end{cases}
    \end{equation*}
\end{corollary}

Also, Theorem \ref{thm: merge-convertible: lower bound} can cover the known bound \cite[Thmeorem 10]{Maturana2020} by setting $r_{F_1}=r_{F_2}=\cdots=r_{F_{t_2}}$.

\begin{corollary}[\cite{Maturana2020},Theorem 10]\label{cor: merge-convertible 2}
    For all $(t_1,1)_q$  MDS generalized merge-convertible codes such that $r_{I_1}=r_{I_2}=\cdots=r_{I_{t_1}}$, the access cost $\rho$ satisfies
    \begin{equation*}
        \rho \ge \begin{cases}
                   \sum_{r_{F_1} \le k_{I_i}}r_{F_1} + \sum_{r_{F_1} > k_{I_i}}k_{I_i}+r_{F_1}, & \mbox{if } r_{F_1} \le r_{I_1}, \\
                   \sum_{i\in[t_1]} k_{I_i}+r_{F_1}, & \mbox{otherwise}.
                 \end{cases}
    \end{equation*}
\end{corollary}

\subsection{Split Regime}

In the split regime, the $t_2$ final codes defined by \eqref{eqn_5} turns into
\begin{equation*}
\mathcal{C}_{F_j}
=\left\{ (\bm{x}_1,\sigma_j(\bm{x}_2)) : \bm{x}\in\mathcal{C}_{I_1}, \bm{x}_1=\bm{x}|_{\unchanged_{1,j}}, \bm{x}_2=\bm{x}|_{\reading_{1,j}} \right\}, \quad j\in[t_2]
\end{equation*}
with $k_{I_1}=\sum_{j\in[t_2]}k_{F_j}$ and the cardinality of the written symbols is
\begin{equation}\label{eqn_write_cost_split}
|\written_j|=n_{F_j}-|\unchanged_{1,j}|.
\end{equation}

We obtain a lower bound on the access cost of  MDS generalized split-convertible codes as follows.

\begin{theorem}\label{thm 1}
    Let $\bm{\mathcal{C}}$ be a  $(1,t_2)_q$  MDS generalized split-convertible code, then
    \begin{equation}\label{eqn_80}
        \left|\bigcup_{j\in[t_2]}\reading_{1,j}\right|
        \ge k_{I_1}
        - \max\left\{ 0, \underset{\substack{j\in[t_2], r_{F_j}\le\\ \min\{k_{F_j},r_{I_1}\} } }{\max}\{k_{F_j}-r_{F_j}\} \right\}.
    \end{equation}
    That is, the access cost $\rho$ satisfies
    \begin{equation}\label{eqn_split_bound}
        \rho
        \ge k_{I_1}
        - \max\left\{ 0, \underset{\substack{j\in[t_2], r_{F_j}\le\\ \min\{k_{F_j},r_{I_1}\} } }{\max}\{k_{F_j}-r_{F_j}\} \right\}
        + \sum_{j \in [t_2]}r_{F_j},
    \end{equation}
    where the equality is achieved only if the equality in \eqref{eqn_80} is attained and $|\unchanged_{1,j}|=k_{F_j}$ for all $j\in[t_2]$.
\end{theorem}

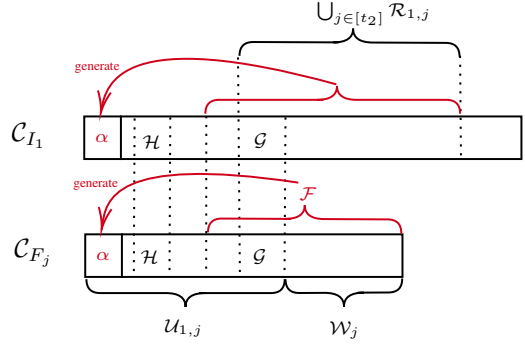
\begin{figure}[t!]
\centering
\tikzset{every picture/.style={line width=0.75pt}} 
\begin{tikzpicture}[x=0.75pt,y=0.75pt,yscale=-1,xscale=1]
\draw   (60.6,69.95) -- (281.1,69.95) -- (281.1,91.45) -- (60.6,91.45) -- cycle ;
\draw   (61.1,129.45) -- (221.1,129.45) -- (221.1,150.95) -- (61.1,150.95) -- cycle ;
\draw  [dash pattern={on 0.84pt off 2.51pt}]  (162.1,70.45) -- (161.6,151.45) ;
\draw   (61.7,152.1) .. controls (61.7,156.77) and (64.03,159.1) .. (68.7,159.1) -- (101.7,159.1) .. controls (108.37,159.1) and (111.7,161.43) .. (111.7,166.1) .. controls (111.7,161.43) and (115.03,159.1) .. (121.7,159.1)(118.7,159.1) -- (154.7,159.1) .. controls (159.37,159.1) and (161.7,156.77) .. (161.7,152.1) ;
\draw   (249.9,39.85) .. controls (249.92,35.18) and (247.6,32.84) .. (242.93,32.82) -- (206.18,32.65) .. controls (199.51,32.62) and (196.19,30.28) .. (196.21,25.61) .. controls (196.19,30.28) and (192.85,32.59) .. (186.18,32.56)(189.18,32.57) -- (145.93,32.38) .. controls (141.26,32.36) and (138.92,34.68) .. (138.9,39.35) ;
\draw  [dash pattern={on 0.84pt off 2.51pt}]  (138.16,40.16) -- (139.1,150.45) ;
\draw    (79.1,69.5) -- (79.1,91.95) ;
\draw    (79.6,129) -- (79.6,151.45) ;
\draw  [color={rgb, 255:red, 208; green, 2; blue, 27 }  ,draw opacity=1 ] (220.5,128.95) .. controls (220.53,124.28) and (218.21,121.94) .. (213.54,121.91) -- (181.79,121.75) .. controls (175.12,121.72) and (171.8,119.37) .. (171.82,114.7) .. controls (171.8,119.37) and (168.46,121.68) .. (161.79,121.65)(164.79,121.67) -- (130.04,121.49) .. controls (125.37,121.47) and (123.03,123.79) .. (123,128.46) ;
\draw [color={rgb, 255:red, 0; green, 0; blue, 0 }  ,draw opacity=1 ] [dash pattern={on 0.84pt off 2.51pt}]  (122.16,69.76) -- (122.1,149.95) ;
\draw  [dash pattern={on 0.84pt off 2.51pt}]  (250.16,39.36) -- (250.6,90.95) ;
\draw  [dash pattern={on 0.84pt off 2.51pt}]  (86,70.95) -- (85.6,149.95) ;
\draw  [dash pattern={on 0.84pt off 2.51pt}]  (104,70.45) -- (103.6,149.45) ;
\draw [color={rgb, 255:red, 208; green, 2; blue, 27 }  ,draw opacity=1 ]   (168.56,103.36) .. controls (74.7,89.3) and (71.45,112.42) .. (69.65,125.06) ;
\draw [shift={(69.36,126.96)}, rotate = 279.46] [color={rgb, 255:red, 208; green, 2; blue, 27 }  ,draw opacity=1 ][line width=0.75]    (10.93,-3.29) .. controls (6.95,-1.4) and (3.31,-0.3) .. (0,0) .. controls (3.31,0.3) and (6.95,1.4) .. (10.93,3.29)   ;
\draw  [color={rgb, 255:red, 208; green, 2; blue, 27 }  ,draw opacity=1 ] (250.16,68.56) .. controls (250.16,63.89) and (247.83,61.56) .. (243.16,61.56) -- (198.08,61.56) .. controls (191.41,61.56) and (188.08,59.23) .. (188.08,54.56) .. controls (188.08,59.23) and (184.75,61.56) .. (178.08,61.56)(181.08,61.56) -- (128.7,61.56) .. controls (124.03,61.56) and (121.7,63.89) .. (121.7,68.56) ;
\draw [color={rgb, 255:red, 208; green, 2; blue, 27 }  ,draw opacity=1 ]   (187.76,53.76) .. controls (80.22,24.12) and (71.59,52.96) .. (69.29,66.21) ;
\draw [shift={(68.96,68.16)}, rotate = 279.46] [color={rgb, 255:red, 208; green, 2; blue, 27 }  ,draw opacity=1 ][line width=0.75]    (10.93,-3.29) .. controls (6.95,-1.4) and (3.31,-0.3) .. (0,0) .. controls (3.31,0.3) and (6.95,1.4) .. (10.93,3.29)   ;
\draw   (162.1,152.1) .. controls (162.13,156.77) and (164.47,159.09) .. (169.14,159.06) -- (181.57,158.99) .. controls (188.24,158.95) and (191.58,161.26) .. (191.61,165.93) .. controls (191.58,161.26) and (194.9,158.91) .. (201.57,158.87)(198.57,158.89) -- (214,158.8) .. controls (218.67,158.77) and (220.99,156.43) .. (220.96,151.76) ;

\draw (99.9,170.8) node [anchor=north west][inner sep=0.75pt]  [font=\scriptsize]  {$\mathcal{U}_{1,j}$};
\draw (182.7,171.8) node [anchor=north west][inner sep=0.75pt]  [font=\scriptsize]  {$\mathcal{W}_{j}$};
\draw (176.2,10.5) node [anchor=north west][inner sep=0.75pt]  [font=\scriptsize]  {$\bigcup _{j\in [ t_{2}]}\mathcal{R}_{1,j}$};
\draw (144.6,76.9) node [anchor=north west][inner sep=0.75pt]  [font=\scriptsize]  {$\mathcal{G}$};
\draw (144.1,135.4) node [anchor=north west][inner sep=0.75pt]  [font=\scriptsize]  {$\mathcal{G}$};
\draw (64.6,77.4) node [anchor=north west][inner sep=0.75pt]  [font=\scriptsize,color={rgb, 255:red, 208; green, 2; blue, 27 }  ,opacity=1 ]  {$\alpha $};
\draw (65.1,136.9) node [anchor=north west][inner sep=0.75pt]  [font=\scriptsize,color={rgb, 255:red, 208; green, 2; blue, 27 }  ,opacity=1 ]  {$\alpha $};
\draw (167.6,104.4) node [anchor=north west][inner sep=0.75pt]  [font=\scriptsize,color={rgb, 255:red, 208; green, 2; blue, 27 }  ,opacity=1 ]  {$\mathcal{F}$};
\draw (22.5,71.9) node [anchor=north west][inner sep=0.75pt]    {$\mathcal{C}_{I_{1}}$};
\draw (24,130.4) node [anchor=north west][inner sep=0.75pt]    {$\mathcal{C}_{F_{j}}$};
\draw (88.5,77.9) node [anchor=north west][inner sep=0.75pt]  [font=\scriptsize]  {$\mathcal{H}$};
\draw (87.5,135.9) node [anchor=north west][inner sep=0.75pt]  [font=\scriptsize]  {$\mathcal{H}$};
\draw (53.6,100.4) node [anchor=north west][inner sep=0.75pt]  [font=\scriptsize,color={rgb, 255:red, 208; green, 2; blue, 27 }  ,opacity=1 ] [align=left] {{\tiny generate}};
\draw (54.4,41.2) node [anchor=north west][inner sep=0.75pt]  [font=\scriptsize,color={rgb, 255:red, 208; green, 2; blue, 27 }  ,opacity=1 ] [align=left] {{\tiny generate}};
\end{tikzpicture}

\caption{The relationship among code symbols used in the proof of Theorem \ref{thm 1}, where $\alpha\in\unchanged_{1,j}\setminus(\bigcup_{j\in[t_2]}\reading_{1,j})$, and $\mathcal{F}\subseteq\unchanged_{1,j}\cup\written_j$ such that $|\mathcal{F}|\ge k_{F_j}, \written_j\subseteq\mathcal{F}, \alpha\not\in\mathcal{F}$, and $\mathcal{G}=\unchanged_{1,j}\cap(\bigcup_{j\in[t_2]}\reading_{1,j})$, and $\mathcal{H}\subseteq\unchanged_{1,j}\setminus\mathcal{G}$ such that $|\mathcal{H}|=r_{I_1}+1$.}
\label{fig:enter-label}
\end{figure}

\begin{IEEEproof}
First of all, we introduce some sets of code symbols for $(1,t_2)_q$  MDS generalized split-convertible code $\bm{\mathcal{C}}$, i.e., for a $j\in[t_2]$,
\begin{align*}
\mathcal{G} &\subseteq\unchanged_{1,j}, \\
\mathcal{F} &=\mathcal{G}\cup\written_j, \\
\mathcal{H} &\subseteq\unchanged_{1,j}\setminus\mathcal{G}.
\end{align*}
Figure \ref{fig:enter-label} intuitively shows the relationship among the above code symbols.

Next, we prove
\begin{equation}\label{Eqn_ac_bound}
 \rho_r= \left|\bigcup_{j\in[t_2]}\reading_{1,j}\right| \ge \min\left\{k_{I_1}, \underset{j\in[t_2]}{\min}\{k_{I_1}+r_{F_j}-|\unchanged_{1,j}|\}\right\}
\end{equation}
for given $|\unchanged_{1,1}|,|\unchanged_{1,2}|,\cdots,|\unchanged_{1,t_2}|$.
In fact, there are two different methods of selecting unchanged symbols and read symbols that would affect the read access cost.

\textbf{Method 1:}  $\unchanged_{1,j}\subseteq\bigcup_{j\in[t_2]}\reading_{1,j}$ for all $j\in[t_2]$.  In this case, $\mathcal{C}_{F_j}=\mathcal{C}_{F_j}|_{\unchanged_{1,j}\cup\written_{j}}$ can be generated by $\mathcal{C}_{I_1}|_{\bigcup_{j\in[t_2]}\reading_{1,j}}$ for all $j\in[t_2]$ since $\mathcal{C}_{F_j}|_{\written_{j}}$ is generated from $\mathcal{C}_{I_1}|_{\reading_{1,j}}$ during the conversion procedure and
$\mathcal{C}_{F_j}|_{\unchanged_{1,j}}=\mathcal{C}_{I_1}|_{\unchanged_{1,j}}$ from \eqref{eqn_1}.
Then, $\dim(\mathcal{C}_{I_1}|_{\bigcup_{j\in[t_2]}\reading_{1,j}})\ge k_{I_1}$ and thus $\rho_r=|\bigcup_{j\in[t_2]}\reading_{1,j}|\ge k_{I_1}$.

\textbf{Method 2:} $\unchanged_{1,j}\nsubseteq \bigcup_{j\in[t_2]}\reading_{1,j}$ for a $j\in[t_2]$. In this case, there exists $\alpha\in\unchanged_{1,j}\setminus\bigcup_{j\in[t_2]}\reading_{1,j}$.
Based on the fact $|\unchanged_{1,j}\cup\written_j|=n_{F_j}> k_{F_j}$, specifically choose $\mathcal{F}$ with $|\mathcal{F}|= k_{F_j}$ such that $\alpha\not\in\mathcal{F}$.

Then, $\mathcal{C}_{F_j}|_{\{\alpha\}}$ can be generated from $\mathcal{C}_{F_j}|_{\mathcal{F}}$ since $\mathcal{C}_{F_j}$ is an MDS code with dimension $k_{F_j}$. Moreover, since
$\mathcal{F} = (\mathcal{F}\cap \unchanged_{1,j})\cup  \written_{j}$ by $\mathcal{C}_{F_j}=\mathcal{C}_{F_j}|_{\unchanged_{1,j}\cup\written_{j}}$, and
$\mathcal{C}_{F_j}|_{\written_j}$ is generated from $\mathcal{C}_{I_1}|_{\reading_{1,j}}$ during the conversion procedure, we have that $\mathcal{C}_{I_1}|_{\{\alpha\}}=\mathcal{C}_{F_j}|_{\{\alpha\}}$ can be generated by $\mathcal{C}_{I_1}|_{(\mathcal{F}\cap\unchanged_{1,j})\cup\bigcup_{j\in[t_2]}\reading_{1,j}}$, which implies
\begin{equation}\label{eqn_50}
    \left|(\mathcal{F}\cap\unchanged_{1,j}) \cup\left(\bigcup_{j\in[t_2]}\reading_{1,j}\right)\right|
    \ge k_{I_1}
\end{equation}
by Lemma \ref{lem:fact}. By the definition of $\mathcal{F}$, we obtain
\begin{align*}
    \left|(\mathcal{F}\cap\unchanged_{1,j}) \cup\left(\bigcup_{j\in[t_2]}\reading_{1,j}\right)\right|
    &\le |\mathcal{F}\cap\unchanged_{1,j}|+\left|\bigcup_{j\in[t_2]}\reading_{1,j}\right|\\
    &=|\mathcal{F}|-|\written_j|+\rho_r\\
    &=k_{F_j}-(n_{F_j}-|\unchanged_{1,j}|)+\rho_r.
\end{align*}
Then,
\begin{align}\label{eqn_83}
    \rho_r
    &\ge k_{I_1}-k_{F_j}+n_{F_j}-|\unchanged_{1,j}|\notag\\
    &= k_{I_1}+r_{F_j}-|\unchanged_{1,j}|.
\end{align}

Therefore,  given size $|\unchanged_{1,j}|$ ($j\in[t_2]$), we can choose one of the above two methods to derive \eqref{Eqn_ac_bound}.

Thirdly, to derive \eqref{eqn_80}, we show that for given $|\unchanged_{1,1}|,|\unchanged_{1,2}|,\cdots,|\unchanged_{1,t_2}|$, the read access cost $\rho_r$ satisfies
\begin{equation}\label{Eqn_SC}
\rho_r=\left|\bigcup_{j\in[t_2]}\reading_{1,j}\right|\ge
\min\left\{ k_{I_1}, \underset{\substack{j\in[t_2], r_{F_j}\le\\ \min\{|\unchanged_{1,j}|,r_{I_1}\} } }{\min}\{k_{I_1}+r_{F_j}-|\unchanged_{1,j}|\} \right\} .
\end{equation}
We then need to determine the exact value of  $\min_{j\in[t_2]}  \{k_{I_1},k_{I_1}+r_{F_j}-|\unchanged_{1,j}|\}$ in \eqref{Eqn_ac_bound} in different cases.

\textbf{Case 1:} $r_{F_j}\le\min\{|\unchanged_{1,j}|,r_{I_1}\}$ for a $j\in[t_2]$.  In this case, $k_{I_1}+r_{F_j}-|\unchanged_{1,j}|\le k_{I_1}$. Then, $\rho_r\ge k_{I_1}+r_{F_j}-|\unchanged_{1,j}|$.

\textbf{Case 2:} $r_{F_j}> |\unchanged_{1,j}|$ for all $j\in[t_2]$. Since $k_{I_1}+r_{F_j}-|\unchanged_{1,j}| > k_{I_1}$ for all $j\in[t_2]$ in this case, we have $\rho_r \ge k_{I_1}$.

\textbf{Case 3:} $|\unchanged_{1,j}|\ge r_{F_j}>r_{I_1}$ for  a $j\in[t_2]$.  In this case, specifically define $\mathcal{G}=\unchanged_{1,j}\cap(\bigcup_{i\in[t_2]}\reading_{1,i})$. We claim that $|\mathcal{F}|\ge k_{F_j}$, which is equivalent to  $|\mathcal{G}|\ge k_{F_j}-|\written_j|= |\unchanged_{1,j}|-r_{F_j}$. Assume on the contrary that $|\mathcal{G}|< |\unchanged_{1,j}|-r_{F_j}$,
which gives
\begin{equation*}
    |\unchanged_{1,j}\setminus\mathcal{G}| = |\unchanged_{1,j}|-|\mathcal{G}|>r_{F_j}>r_{I_1}.
\end{equation*}
Recall that $\mathcal{H}\subseteq\unchanged_{1,j}\setminus\mathcal{G}$. Especially, pick  $\mathcal{H}$ with $|\mathcal{H}|=r_{I_1}+1$. Then,
\begin{equation*}
    |\written_j|+|\unchanged_{1,j}\setminus\mathcal{H}|= n_{F_j}-|\mathcal{H}|=  n_{F_j}-(r_{I_1}+1)\ge n_{F_j}-r_{F_j} =k_{F_j},
\end{equation*}
which implies that $\mathcal{C}_{F_j}|_{\mathcal{H}}$ can be generated from $\mathcal{C}_{F_j}|_{(\unchanged_{1,j}\setminus\mathcal{H})\cup\written_j}
\subseteq\mathcal{C}_{I_1}|_{\unchanged_{1,j}\setminus\mathcal{H}}\otimes\mathcal{C}_{F_j}|_{\written_j}$ since $\mathcal{C}_{F_j}$ is an MDS code with dimension $k_{F_j}$. Recall from the conversion procedure that $\mathcal{C}_{F_j}|_{\written_j}$ is generated from
$\mathcal{C}_{I_1}|_{\reading_{1,j}}$.
Hence, $\mathcal{C}_{I_1}|_{\mathcal{H}}$ can be generated by $\mathcal{C}_{I_1}|_{(\unchanged_{1,j}\setminus\mathcal{H})\cup(\bigcup_{j\in[t_2]}\reading_{1,j})}$.
Note that herein we utilize \eqref{eqn_1}, i.e., $\mathcal{C}_{F_j}|_{\unchanged_{1,j}\setminus\mathcal{H}}
=\mathcal{C}_{I_1}|_{\unchanged_{1,j}\setminus\mathcal{H}}$ and $\mathcal{C}_{I_1}|_{\mathcal{H}}=\mathcal{C}_{F_j}|_{\mathcal{H}}$ due to $\mathcal{H}\subseteq\unchanged_{1,j}$.
By $$\mathcal{H}\subseteq\unchanged_{1,j}\setminus\mathcal{G}=\unchanged_{1,j}\setminus \left(\bigcup_{j\in[t_2]}\reading_{1,j}\right),$$  we have
\begin{equation*}
    \mathcal{H} \cap \left((\unchanged_{1,j}\setminus\mathcal{H})\cup\left(\bigcup_{j\in[t_2]}\reading_{1,j}\right)\right) = \emptyset.
\end{equation*}
According to Lemma \ref{lem:fact},
\begin{equation*}
\left|(\unchanged_{1,j}\setminus\mathcal{H})\cup\left(\bigcup_{j\in[t_2]}\reading_{1,j}\right)\right| \ge k_{I_1}.
\end{equation*}
Then, we get a contradiction, i.e.,
\begin{equation*}
    n_{I_1} \ge |\mathcal{H}| + \left|(\unchanged_{1,j}\setminus\mathcal{H})\cup\left(\bigcup_{j\in[t_2]}\reading_{1,j}\right)\right| \ge r_{I_1}+1+k_{I_1} = n_{I_1}+1.
\end{equation*}

For this set $\mathcal{F}$ with size  $|\mathcal{F}|\ge k_{F_j}$, in this case, we can further improve the lower bound on
access cost of method 2 from  $k_{I_1}+r_{F_j}-|\unchanged_{1,j}|$ to  $\rho_r\ge k_{I_1}$. Actually, it follows from $\mathcal{F}=\mathcal{G}\cup\mathcal{W}_j$ and $\mathcal{G}=\unchanged_{1,j}\cap(\bigcup_{j\in[t_2]}\reading_{1,j})$ that
\begin{equation*}
\mathcal{F}\cap\unchanged_{1,j}
=(\mathcal{G}\cap\unchanged_{1,j}) \cup (\written_j\cap\unchanged_{1,j})
=\mathcal{G}\cap\unchanged_{1,j}
\subseteq\bigcup_{j\in[t_2]}\reading_{1,j}.
\end{equation*}
Then, we deduce $\rho_r=|\bigcup_{j\in[t_2]}\reading_{1,j}|\ge k_{I_1}$ from \eqref{eqn_50}.

Combining the above three cases, we obtain the inequality \eqref{Eqn_SC}. According to Lemma \ref{prop_1}, $|\unchanged_{1,j}|\le k_{F_j}$ for $j\in[t_2]$. Therefore, we arrive at \eqref{eqn_80} and \eqref{eqn_split_bound}  by maximizing each $|\unchanged_{1,j}|$ in \eqref{Eqn_SC} and \eqref{eqn_write_cost_split}, i.e., setting $|\unchanged_{1,j}|=k_{F_j}$ for $j\in[t_2]$. This is to say that the lower bound given by \eqref{eqn_80} and \eqref{eqn_split_bound} can be achieved with equality only if $|\unchanged_{1,j}|=k_{F_j}$ for $j\in[t_2]$.
\end{IEEEproof}

\begin{definition}
A $(1,t_2)_q$  MDS generalized split-convertible code $\bm{\mathcal{C}}$ is said to be access-optimal if the read access cost and access cost respectively achieve the lower bounds with equality in \eqref{eqn_80} and \eqref{eqn_split_bound}.
\end{definition}

The following corollary is straightforward from Theorem \ref{thm 1}
and Definition \ref{def: stable}.

\begin{corollary}
    All $(1,t_2)_q$ access-optimal MDS generalized split-convertible codes are stable.
\end{corollary}

Theorem \ref{thm 1} can cover the known bound \cite[Theorem 6]{Maturana2020} by setting $\mathcal{C}_{F_1}=\mathcal{C}_{F_2}=\cdots=\mathcal{C}_{F_{t_2}}$.

\begin{corollary}[{\cite[Theorem 6]{Maturana2020}}]
    For all  $(1,t_2)_q$ MDS generalized split-convertible codes such that $\mathcal{C}_{F_1}=\mathcal{C}_{F_2}=\cdots=\mathcal{C}_{F_{t_2}}$, the access cost $\rho$ satisfies
    \begin{equation*}
        \rho \ge \begin{cases}
                   (t_2-1)k_{F_1}+\min\{r_{F_1},k_{F_1}\}+t_2 r_{F_1}, & \mbox{if } r_{I_1}>r_{F_1}\\
                   t_2 n_{F_1}, & \mbox{otherwise}.
                 \end{cases}
    \end{equation*}
\end{corollary}

Also, Theorem \ref{thm 1} can cover the known bound \cite[Thmeorem 9]{Maturana2020} by setting $r_{F_1}=r_{F_2}=\cdots=r_{F_{t_2}}$.

\begin{corollary}[{\cite[Theorem 9]{Maturana2020}}]
    For all $(1,t_2)_q$  MDS generalized split-convertible codes such that $r_{F_1}=r_{F_2}=\cdots=r_{F_{t_2}}$, the access cost $\rho$ satisfies
    \begin{equation*}
        \rho \ge k_{I_1}-\max\left\{0, \max_{\substack{ j\in[t_2]\\ r_{F_1}\le r_{I_1}}}\{k_{F_j}\}-r_{F_1}\right\}+t_2 r_{F_1}.
    \end{equation*}
\end{corollary}

According to Theorem \ref{thm 1}, a $(1,t_2)_q$ MDS generalized  split-convertible code $\bm{\mathcal{C}}$ is access-optimal if
only one $j\in [t_2]$ final code $\mathcal{C}_{F_j}$ needs to read
$k_{I_1}+r_{F_j}-k_{F_j}<k_{I_1}$  symbols from initial code $\mathcal{C}_{I_1}$. This requirement can be easily achieved as follows.

Firstly, the conversion procedure for final codes $\mathcal{C}_{F_i}\ (i\in[t_2]\setminus\{j\})$ is trivial by reading $k_{F_i}$ unchanged symbols based on default approach. Next, as for the final code $\mathcal{C}_{F_j}$ satisfying $r_{F_j}\le\min\{ k_{F_j}, r_{I_1} \}$, let $\mathcal{V}_j\subseteq(\{1\}\times[n_{I_1}])\setminus\bigcup_{j\in[t_2]}\unchanged_{1,j}$ with $|\mathcal{V}_j|=r_{F_j}$ and $|\unchanged_{1,j}|=k_{F_j}$. According to Lemma \ref{lem: prop of puc and srt}, the punctured code $\mathcal{C}_{I_1}|_{\bigcup_{j\in[t_2]}\unchanged_{1,j}\cup\mathcal{V}_j}$ is a $[k_{I_1}+r_{F_j},k_{I_1}]_q$ MDS code. Suppose that  $\overline{H}_{I_1}$ is its parity check matrix.  Generate the final code $\mathcal{C}_{F_j}$ by means of the parity check matrix $\overline{H}_{I_1}|_{\unchanged_{1,j}\cup\mathcal{V}_j}$. Since any $r_{F_j}$ columns of $\overline{H}_{I_1}$ are linear independent over $\mathbb{F}_q$,  $\mathcal{C}_{F_j}$ is an $[n_{F_j},k_{F_j}]_q$ MDS code. Note that
\begin{align*}
\bm{c}_{F_j}|_{\written_{j}}(\overline{H}_{I_1}|_{\mathcal{V}_j})^\top
&= -\bm{c}_{F_j}|_{\unchanged_{1,j}}(\overline{H}_{I_1}|_{\unchanged_{1,j}})^\top\\
&= -\bm{c}_{I_1}|_{\unchanged_{1,j}}(\overline{H}_{I_1}|_{\unchanged_{1,j}})^\top\\
&= \bm{c}_{I_1}|_{\bigcup_{i\in[t_2]\setminus\{j\}}\unchanged_{1,i}\cup\mathcal{V}_j}(\overline{H}_{I_1}|_{\bigcup_{i\in[t_2]\setminus\{j\}}\unchanged_{1,i}\cup\mathcal{V}_j})^\top
\end{align*}
where $\bm{c}_{I_1}=(\bm{c}_{F_1}|_{\unchanged_{1,1}},\cdots,\bm{c}_{F_{t_2}}|_{\unchanged_{1,t_2}},c_{1,k_{I_1}+1},\cdots,c_{1,n_{I_1}})\in\mathcal{C}_{I_1}$ and $\bm{c}_{F_\ell}\in\mathcal{C}_{F_\ell}$ for $\ell\in[t_2]$. That is, the read access cost is $k_{I_1}+r_{F_j}-k_{F_j}<k_{I_1}$.

Therefore, we concentrate on access-optimal  MDS generalized merge-convertible codes in the last part of this paper.

\section{A characterization of access-optimal merge-convertible codes}\label{sec-structure}
In this section, we present a characterization for access-optimal  MDS generalized merge-convertible codes, focusing on the perspective of parity check matrices. This characterization offers insights into access-optimal  MDS generalized merge-convertible codes.
To begin with, we prove two useful lemmas.

\begin{lemma}\label{lem: bijiao}
    Let $\bm{\mathcal{C}}$ be a $(t_1,1)_q$  access-optimal  MDS generalized merge-convertible code. For $i\in[t_1]$, if the initial code $\mathcal{C}_{I_i}$ satisfies $r_{F_1}<k_{I_i}$ and $r_{F_1}\le r_{I_i}$, then $\reading_{i,1} \cap \unchanged_{i,1} = \emptyset$.
\end{lemma}

\begin{IEEEproof}
   Without loss of generality, let $i=1$. Assume on the contrary that $\reading_{1,1} \cap \unchanged_{1,1} \ne \emptyset$.
     Define $t_1$ initial codes as
     \begin{equation}\label{Eqn_Ci_New}
        \mathcal{C}_{I_i}' =
        \begin{cases}        \mathcal{C}_{I_1}|_{\unchanged_{1,1}\cup\reading_{1,1}}, & \mbox{if } i=1,\\
            \mathcal{C}_{I_i}, & \mbox{if } i\in[t_1]\setminus\{1\}
        \end{cases}
    \end{equation}
    with the  unchanged symbols set $\unchanged_{i,1}'=\unchanged_{i,1}$ and reading symbols set $\reading_{i,1}'=\reading_{i,1}$ for all $i\in [t_1]$.
    Applying the conversion procedure function $\sigma_1$ of original code $\bm{\mathcal{C}}$ to these $t_1$ new initial codes, we can get a new generalized merge-convertible code $\bm{\mathcal{C}}'$.

    Obviously,
    $\mathcal{C}_{F_1}'=\mathcal{C}_{F_1}$ by
    substituting \eqref{Eqn_Ci_New} into  \eqref{Eqn_Gen_Fj}
    and \eqref{eqn_5}, i.e., $\mathcal{C}_{F_1}'$ is an $[n_{F_1},k_{F_1}]_q$ MDS code.
    Then, by \eqref{equ:dimension} we have $\dim(\mathcal{C}_{I_1}|_{\unchanged_{1,1}\cup\reading_{1,1}})=k_{I_1}$, which together with Lemma \ref{lem: prop of puc and srt} indicate that
   $\mathcal{C}_{I_1}|_{\unchanged_{1,1} \cup \reading_{1,1}}$ is  an $[n_{I_1}',k_{I_1}]_{q}$ MDS code where $n_{I_1}'=|\unchanged_{1,1}\cup\reading_{1,1}|$.
  Hence, $\bm{\mathcal{C}}'$ is an  MDS generalized merge-convertible code.

Recall that $\bm{\mathcal{C}}$ is access-optimal MDS generalized merge-convertible code and $r_{F_1} \le \min\{ r_{I_1},k_{I_1} \}$, thus $|\unchanged_{1,1}| = k_{I_1}$ and  $|\reading_{1,1}| = r_{F_1}$ by Theorem \ref{thm: merge-convertible: lower bound} and Definition \ref{Def_access_optimal}, respectively.
Then,
    \begin{equation*}
        n_{I_1}' <  |\unchanged_{1,1}|+|\reading_{1,1}| =k_{I_1}+r_{F_1},
    \end{equation*}
which results in
    \begin{equation*}\label{equ: lem bujiao 1}
        r_{I_1}'=n_{I_1}' - k_{I_1} < r_{F_1}.
    \end{equation*}
Hence, it follows from Theorem \ref{thm: merge-convertible: lower bound} that $k_{I_1}\le|\reading_{1,1}'|$. However, $|\reading_{1,1}'|=|\reading_{1,1}|=r_{F_1}<k_{I_1}$, a contradiction, which completes the proof.
\end{IEEEproof}

We introduce some commonly used notation below. Let $\bm{\mathcal{C}}$ be a $(t_1,1)_q$ generalized merge-convertible code. Let $G_{I_i}$ and $H_{I_i}$ be generator and parity check matrices of the initial code $\mathcal{C}_{I_i}$ for $i\in[t_1]$, respectively. And denote generator and parity check matrices of the final code $\mathcal{C}_{F_1}$ as $G_{F_1}$ and $H_{F_1}$, respectively.

\begin{lemma}\label{lem_21}
Let $\bm{\mathcal{C}}$ be a linear $(t_1,1)_q$  access-optimal MDS generalized merge-convertible code.
For $i\in[t_1]$, if the initial code $\mathcal{C}_{I_i}$ satisfies $r_{F_1}<k_{I_i}$ and $r_{F_1}\le r_{I_i}$, then there exists a parity check matrix $\overline{H}_{I_i}$ for punctured code $\mathcal{C}_{I_i}|_{\unchanged_{i,1}\cup\reading_{i,1}}$ such that
$H_{F_1}|_{\unchanged_{i,1}} = \overline{H}_{I_i}|_{\unchanged_{i,1}}
$.
\end{lemma}

\begin{IEEEproof}
Without loss of generality, set $i=1$. Let $\bm{c}_{F_1}=(\bm{c}_{I_1}|_{\unchanged_{1,1}},\bm{0}|_{\unchanged_{2,1}},\cdots,\bm{0}|_{\unchanged_{t_1,1}},\bm{c}_{F_1}|_{\written_1})$ be a codeword of the final code $\mathcal{C}_{F_1}$. Then,
\begin{equation*}
\bm{c}_{F_1}\cdot H_{F_1}^\top
=(\bm{c}_{I_1}|_{\unchanged_{1,1}},\bm{0}|_{\unchanged_{2,1}},\cdots,\bm{0}|_{\unchanged_{t_1,1}},\bm{c}_{F_1}|_{\written_1})
\cdot
(H_{F_1}|_{\unchanged_{1,1}},\cdots,H_{F_1}|_{\unchanged_{t_1,1}},H_{F_1}|_{\written_1})^\top
=\bm{0},
\end{equation*}
which implies
\begin{equation}\label{eqn_51}
    \bm{c}_{I_1}|_{\unchanged_{1,1}} (H_{F_1}|_{\unchanged_{1,1}})^\top
    +\bm{c}_{F_1}|_{\written_1}(H_{F_1}|_{\written_1})^\top  =\bm{0}.
\end{equation}
Since $\bm{\mathcal{C}}$ is access-optimal, then $|\unchanged_{i,1}|=k_{I_i}$ for all $i\in[t_1]$ by Theorem \ref{thm: merge-convertible: lower bound}.
Then, $\mathcal{C}_{I_i}|_{\reading_{i,1}}$ can be generated from $\mathcal{C}_{I_i}|_{\unchanged_{i,1}}$ for $i\in[t_1]$ since $\mathcal{C}_{I_i}$ is an MDS code with dimension $k_{I_i}$. Recall from conversion procedure that $\mathcal{C}_{F_1}|_{\written_1}$ is generated from
$\bigotimes_{i\in[t_1]}\mathcal{C}_{I_i}|_{\reading_{i,1}}$.
Hence, $\mathcal{C}_{F_1}|_{\written_1}$ can be generated by $\bigotimes_{i\in[t_1]\setminus\{1\}}\mathcal{C}_{I_i}|_{\unchanged_{i,1}}\otimes\mathcal{C}_{I_1}|_{\reading_{1,1}}$.  Because $\bm{\mathcal{C}}$ is linear,  the conversion  function $\sigma_1$ is a linear function determined by a $(r_{F_1}+k_{F_1}-k_{I_1})\times r_{F_1}$ matrix $Q^*$ such that
\begin{equation}\label{eqn_52}
\bm{c}_{F_1}|_{\written_1}
=(\bm{c}_{I_1}|_{\reading_{1,1}},\bm{0}|_{\unchanged_{2,1}},\cdots,\bm{0}|_{\unchanged_{t_1,1}}) Q^*
=\bm{c}_{I_1}|_{\reading_{1,1}} Q,
\end{equation}
where $Q$ is a $r_{F_1}\times r_{F_1}$  submatrix of $Q^*$.

Note that $r_{F_1}<k_{I_1}$ and $r_{F_1}\le r_{I_1}$. Thus, $\unchanged_{1,1}\cap\reading_{1,1}=\emptyset$ by Lemma \ref{lem: bijiao} and $|\reading_{1,1}|=r_{F_1}$ by Theorem \ref{thm: merge-convertible: lower bound}, which indicates that $\mathcal{C}_{I_1}|_{\unchanged_{1,1}\cup\reading_{1,1}}$
is a $[k_{I_1}+r_{F_1},k_{I_1}]_q$ MDS code.
Combining \eqref{eqn_51} with \eqref{eqn_52}, we have
\begin{equation*}
    (\bm{c}_{I_1}|_{\unchanged_{1,1}},\ \bm{c}_{I_1}|_{\reading_{1,1}}) \cdot (H_{F_1}|_{\unchanged_{1,1}},\ H_{F_1}|_{\written_1} Q^\top)^\top =\bm{0}
\end{equation*}
holds for all codewords of $\mathcal{C}_{I_1}|_{\unchanged_{1,1}\cup\reading_{1,1}}$. We conclude that $\rank(H_{F_1}|_{\unchanged_{1,1}})=\min\{r_{F_1},k_{I_1}\}=r_{F_1}$
since $\mathcal{C}_{F_1}$ is an $[n_{F_1},k_{F_1}]$ MDS code. Therefore,  $\overline{H}_{I_1}=(H_{F_1}|_{\unchanged_{1,1}},\ H_{F_1}|_{\written_1} Q^\top)$ is a parity check matrix of punctured code $\mathcal{C}_{I_1}|_{\unchanged_{1,1}\cup\reading_{1,1}}$ with $
\overline{H}_{I_1}|_{\unchanged_{1,1}} = H_{F_1}|_{\unchanged_{1,1}}
$.
\end{IEEEproof}

Now we are ready to present a necessary and sufficient condition for access-optimal  MDS generalized merge-convertible codes in the view of parity check matrices.

\begin{theorem}\label{structure}
For any $(t_1,1)_q$  MDS generalized  merge-convertible code $\bm{\mathcal{C}}$, define
\begin{eqnarray}\label{Eqn_Def_S}
\mathcal{S} \triangleq \{ i: r_{F_1}<k_{I_i}, r_{F_1}\le r_{I_i}, i\in[t_1] \}.
\end{eqnarray}
Then,  $\bm{\mathcal{C}}$ is access-optimal if and only if  $\unchanged_{i,1},\reading_{i,1}\subseteq (\{i\}\times[n_{I_i}])$ with $|\unchanged_{i,1}|=k_{I_i}$ such that
\begin{equation*}
\begin{cases}
|\reading_{i,1}| = r_{F_1}, \unchanged_{i,1}\cap \reading_{i,1}=\emptyset, &\mbox{if } i\in\mathcal{S},\\
|\reading_{i,1}| = k_{I_i}, &\mbox{if } i\in[t_1]\setminus\mathcal{S},
\end{cases}
\end{equation*}
and a parity check matrix $\overline{H}_{I_i}$ for punctured code $\mathcal{C}_{I_i}|_{\unchanged_{i,1}\cup\reading_{i,1}}$ for $i\in\mathcal{S}$ satisfies
\begin{equation}\label{eqn_43}
H_{F_1}|_{\unchanged_{i,1}} = \overline{H}_{I_i}|_{\unchanged_{i,1}}.
\end{equation}
\end{theorem}

\begin{IEEEproof}
Since necessity directly follows from Lemma \ref{lem: bijiao} and Lemma \ref{lem_21}, we only show the sufficiency herein.

Assume that a codeword $\bm{c}_{F_1}=(\bm{c}_{I_1}|_{\unchanged_{1,1}},\cdots,\bm{c}_{I_{t_1}}|_{\unchanged_{t_1,1}},\bm{c}_{F_1}|_{\written_{1}})\in \mathcal{C}_{F_1}$.  The written symbols $\bm{c}_{F_1}|_{\written_1}$ can be generated as
    \begin{align}
        \bm{c}_{F_1}|_{\written_{1}} (H_{F_1}|_{\written_1})^\top &= - \sum_{i\in[t_1]} \bm{c}_{I_i}|_{\unchanged_{i,1}} (H_{F_1}|_{\unchanged_{i,1}})^\top \notag\\
        &= - \sum_{i \in \mathcal{S}} \bm{c}_{I_i}|_{\unchanged_{i,1}}  (\overline{H}_{I_i}|_{\unchanged_{i,1}})^\top  - \sum_{i \in [t_1] \setminus \mathcal{S}} \bm{c}_{I_i}|_{\unchanged_{i,1}} (H_{F_1}|_{\unchanged_{i,1}})^\top \notag\\
        &= \sum_{i\in\mathcal{S}}\bm{c}_{I_i}|_{\reading_{i,1}} (\overline{H}_{I_i}|_{\reading_{i,1}})^\top - \sum_{i \in [t_1] \setminus \mathcal{S}} \bm{c}_{I_i}|_{\unchanged_{i,1}} (H_{F_1}|_{\unchanged_{i,1}})^\top,\label{eqn_60}
    \end{align}
    where the first equation holds by $\bm{c}_{F_1} \cdot H_{F_1}^\top=\bm{0}$ and $\bm{c}_{F_1}|_{\unchanged_{i,1}}=\bm{c}_{I_i}|_{\unchanged_{i,1}}$, the second equation follows from \eqref{eqn_43}, and the last equation
    is due to
    $\bm{c}_{I_i}|_{\unchanged_{i,1}\cup\reading_{i,1}} \cdot \overline{H}_{I_i}^\top=\bm{0}$ for all $i\in\mathcal{S}$.
    Hence, the read access cost resulting from the above equation is
    \begin{equation*}
       \sum_{i\in\mathcal{S}}|\reading_{i,1}| + \sum_{i\in[t_1]\setminus\mathcal{S}}|\unchanged_{i,1}| =r_{F_1} |\mathcal{S}|  + \sum_{i \in [t_1] \setminus \mathcal{S}} k_{I_i},
    \end{equation*}
    and the corresponding write access cost is $|\written_{1}| = r_{F_1}$, which are optimal with respect to the lower bound in Theorem \ref{thm: merge-convertible: lower bound}.
\end{IEEEproof}

\section{An Explicit Construction: An Application of Theorem \ref{structure}}\label{sec_cons}

Maturana and Rashmi \cite{Maturana2022}, Kong \cite{Kong2023} respectively proposed some constructions of conventional access-optimal  MDS merge-convertible codes corresponding to those in Corollary \ref{cor: merge-convertible 1}.  In fact, their constructions satisfy $\mathcal{C}_{I_1}=\mathcal{C}_{I_2}=\cdots=\mathcal{C}_{I_{t_1}}$ and $r_{F_1}<k_{I_i},r_{F_1}\le r_{I_i}$ for all $i\in[t_1]$.  It is easy to check that their parity check matrices satisfy \eqref{eqn_43} in Theorem \ref{structure}.

Given $t_1$  MDS initial codes $\mathcal{C}_{I_i}$  ($i\in[t_1]$) with  parity check matrix $H_{I_i}$, we can construct a generalized merge-convertible code $\bm{\mathcal{C}}$ from \eqref{eqn_60} by choosing some matrices $H_{F_1}|_{\unchanged_{i,1}}$ for $i\not\in S$. According to Theorem \ref{structure}, $\bm{\mathcal{C}}$ is access-optimal if $\mathcal{C}_{F_1}$ is an MDS code. Therefore, the key is finding suitable parity check matrices such that the resultant code $\mathcal{C}_{F_1}$ is MDS code.

To this end, we employ the well-known Vandermond matrix to construct the parity check matrices in this section. Moreover, according to Definition \ref{def_vand}, the concatenation of Vandermond matrices and an extended Vandermond-type matrix is still an extended Vandermond-type matrix, i.e.,
\begin{equation*}
(\mathbf{Vand}_{r,n}(\gamma,\mathbf{w})|_{[n-1]}, \mathbf{Vand}_{r,m}(\gamma',\mathbf{w}'))
=\mathbf{Vand}_{r,n-1+m}((\gamma,\gamma'),(\mathbf{w}|_{[n-1]},\mathbf{w}'))
\end{equation*}
if $\gamma_i\ne \gamma_j^*$ for $i\in [n-1]$ and $j\in [m-1]$. This is exactly what we need to choose all the parity check matrices, i.e., $H_{I_i}$ for $i\in [t_1]$ and $H_{F_1}$ to be extended Vandermonde-type, and  $H_{F_1}|_{\unchanged_{i,1}}$ for $i\not\in S$ to be Vandermonde-type.

\begin{construction}\label{con: GRS}
For a positive integer $t_1>1$, let $n_{I_i},k_{I_i}$ for $i\in[t_1]$ and $r_{F_1}$ be positive integers, where $k_{I_i}<n_{I_i}$.  Choose $\mathbf{w}_i=(w_{i,1},w_{i,2},\cdots,w_{i,n_{I_i}})$ over $\mathbb{F}_q^*$, and choose $\gamma_i=(\gamma_{i,1},\gamma_{i,2},\cdots,\gamma_{i,n_{I_i}-1})$,  $\gamma'=(\gamma_1',\cdots,\gamma_{r_{F_1}-1}')$ over $\mathbb{F}_q$,  satisfying the following two properties:
\begin{enumerate}
\item [P1.] $\gamma_i$ contains $n_{I_i}-1$ distinct elements for $i\in[t_1]$;
\item [P2.] There exist sets $\unchanged_{i,1}\subseteq\{i\}\times[n_{I_i}-1]$ with $|\unchanged_{i,1}|=k_{I_i}$ for $i\in[t_1]$ such that
\begin{equation}\label{eqn_gamma_star}
\gamma^*\triangleq(\gamma_1|_{\unchanged_{1,1}},\cdots,\gamma_{t_1}|_{\unchanged_{t_1,1}},\gamma')
\end{equation}
contains $n_{F_1}-1=\sum_{i\in[t_1]}k_{I_i}+r_{F_1}-1$ distinct elements.
\end{enumerate}

Firstly, generate $t_1$ initial codes $\mathcal{C}_{I_i}$ by means of the parity check matrix $\mathbf{Vand}_{r_{I_i},n_{I_i}}(\gamma_i,\mathbf{w}_i)$ for $i\in[t_1]$.
Without loss of generality,  write $\mathcal{S}=[\ell]$ in \eqref{Eqn_Def_S}   for some $0\le\ell\le t_1$. Choose any set $\reading_{i,1}\subseteq(\{i\}\times[n_{I_i}])\setminus\unchanged_{i,1}$ with $(i,n_{I_i})\in\reading_{i,1}$ and $|\reading_{i,1}|=r_{F_1}$ for $i\in[\ell]$.
At last, construct the final code $\mathcal{C}_{F_1}$ by the parity check matrix $\mathbf{Vand}_{r_{F_1},n_{F_1}}(\gamma^*,\mathbf{w}^*)$ where
\begin{align*}
\mathbf{w}^* &= (\theta_{\unchanged_{1,1}\cup\reading_{1,1}}(\mathbf{w}_1)|_{\unchanged_{1,1}} ,\cdots, \theta_{\unchanged_{\ell,1}\cup\reading_{\ell,1}}(\mathbf{w}_\ell)|_{\unchanged_{\ell,1}} , \mathbf{w}_{\ell+1}|_{\unchanged_{\ell+1,1}}, \cdots, \mathbf{w}_{t_1}|_{\unchanged_{t_1,1}}, \mathbf{1}_{r_{F_1}}),
\end{align*}
$\theta_{\unchanged_{i,1}\cup\reading_{i,1}}(\mathbf{w}_i)$ for $i\in [\ell]$ is  gievn by Lemma \ref{lem_GRS}, $\gamma^*$ is denfined by \eqref{eqn_gamma_star}, and
$\mathbf{1}_{r_{F_1}}$ is the all-one vector of length $r_{F_1}$.
\end{construction}

\begin{theorem}\label{thm_4}
For a positive integer $t_1>1$, let $n_{I_i},k_{I_i}$ for $i\in[t_1]$ and $r_{F_1}$ be positive integers, where $k_{I_i}<n_{I_i}$. If a prime power $q\ge \max\{n_{I_1},\cdots,n_{I_{t_1}},n_{F_1}\}-1$, a $(t_1,1)_q$   MDS generalized merge-convertible code  with
optimal access cost and parameters $n_{I_i},k_{I_i}$ for $i\in[t_1]$ can be generated from Construction \ref{con: GRS},
where $n_{F_1}=k_{F_1}+r_{F_1}$ and $k_{F_1}=\sum_{i\in[t_1]}k_{I_i}$.
\end{theorem}

\begin{IEEEproof}
When $q\ge \max\{n_{I_1},\cdots,n_{I_{t_1}},n_{F_1}\}-1$, the vectors $\mathbf{w}_i,\gamma_i$ ($i\in[t_1]$), and $\gamma'$ satisfying P1 and P2 is well-defined.
By applying P1 to Lemma \ref{lem_GRS}, $\mathcal{C}_{I_i}$ is $[n_{I_i},k_{I_i}]_q$ MDS code. According to P2,  $\gamma^*$ is an $(n_{F_1}-1)$-tuple of $\sum_{i\in [t_1]}k_{I_i}+r_{F_1}-1$ distinct elements, and $\mathbf{w}^*$ contains $\sum_{i\in [t_1]}k_{I_i}+r_{F_1}$ non-zero elements, together with Lemma \ref{lem_GRS}, which indicated that $\mathcal{C}_{F_1}$ is an $[n_{F_1},k_{F_1}]_q$  MDS code. It then follows from Theorem \ref{structure} that  $\bm{\mathcal{C}}$ is an access-optimal  MDS generalized merge-convertible code.
\end{IEEEproof}

\begin{remark}\label{rem_3}
Compared to constructions in \cite{Maturana2022} and \cite[Corollary II.2]{Kong2023}, Theorem \ref{thm_4} not only relaxes condition $\mathcal{C}_{I_1}=\mathcal{C}_{I_2}=\cdots=\mathcal{C}_{I_{t_1}}$ but also does not relies on a specific structure, for examples, the superregular Hankel arrays in \cite{Maturana2022} or the multiplicative group over $\mathbb{F}_q$ in  \cite{Kong2023}.
The detailed comparison of the parameters is shown in Table \ref{compare}.

Notably, the field size  in Theorem \ref{thm_4} is the theoretical minimum due to the  MDS-Conjecture \cite{Huffman2003}: If there is an $[n,k]_q$ MDS code with $2\le k\le n-2$, then $n\le q+1$, except that when $q$ is even and $k=3$ or $k=q-1$, $n\le q+2$.

\end{remark}

Finally, it should be noted that our construction can employ more kinds of matrices as parity matrices besides the extended Vandermond-type matrix. For example, based on the parity check matrices of Gabidulin codes \cite{Gabidulin1985} or linearized Reed-Solomon codes \cite{Martinez2018}, we can get access-optimal MDS generalized merge-convertible codes as applications of our characterization. Nevertheless, the finite field size is not as good as the extended GRS codes, so we omit herein.

\section{Conclusion}\label{sec_conclusion}
In this paper, the convertible codes were generalized to allow initial and final codes with different parameters. Accordingly, lower bounds on the access cost were established for generalized MDS convertible codes in the merge and split regimes. A characterization of access-optimal MDS generalized merge-convertible codes was introduced via parity check matrices. Finally, as an application of our characterization, a construction of MDS generalized merge-convertible codes with optimal access cost was obtained using extended GRS codes.

As mentioned before, the codes of all the above studies are linear such as almost affine codes. Actually, our lower bounds can be easily extended to non-linear scenarios.
However, the characterization and construction for non-linear cases remain open and are left for further research.

\end{document}